\documentclass[twocolumn,showpacs,superscriptaddress,floatfix,longbibliography]{revtex4}
\usepackage{graphicx,amsfonts,amssymb,amsmath,xspace}
\usepackage[colorlinks=true,citecolor=green,linkcolor=red,urlcolor=blue]{hyperref}
\usepackage{bbm}

\usepackage{color}
\definecolor{g}{rgb}{.1,0.4,.1} 
\definecolor{b}{rgb}{0,0.2,1}
\definecolor{rouge}{rgb}{0.82,0.,0.}
\definecolor{vert}{rgb}{0.,0.82,0.}
\definecolor{orange}{rgb}{1,0.5,0.}
\definecolor{bleu}{rgb}{0.,0.,0.82}
\definecolor{m}{rgb}{0.82,0.,0.82}
\definecolor{vert2}{rgb}{0.,0.5,0.}
\definecolor{rougeclair}{rgb}{1.0,0.7,0.7}

\newcommand{\be}{\begin{equation}}
\newcommand{\ee}{\end{equation}}

\def\nbC{{\mathchoice {\setbox0=\hbox{$\displaystyle\rm C$}%
\hbox{\hbox to0pt{\kern0.4\wd0\vrule height0.9\ht0\hss}\box0}} 
{\setbox0=\hbox{$\textstyle\rm
C$}\hbox{\hbox to0pt{\kern0.4\wd0\vrule height0.9\ht0\hss}\box0}} 
{\setbox0=\hbox{$\scriptstyle\rm
C$}\hbox{\hbox to0pt{\kern0.4\wd0\vrule height0.9\ht0\hss}\box0}}
{\setbox0=\hbox{$\scriptscriptstyle\rm C$}\hbox{\hbox to0pt{\kern0.4\wd0\vrule
height0.9\ht0\hss}\box0}}}}
\def\nbQ{{\mathchoice {\setbox0=\hbox{$\displaystyle\rm 
Q$}\hbox{\raise 0.15\ht0\hbox
to0pt{\kern0.4\wd0\vrule height0.8\ht0\hss}\box0}} 
{\setbox0=\hbox{$\textstyle\rm Q$}\hbox{\raise
0.15\ht0\hbox to0pt{\kern0.4\wd0\vrule height0.8\ht0\hss}\box0}} 
{\setbox0=\hbox{$\scriptstyle\rm
Q$}\hbox{\raise 0.15\ht0\hbox to0pt{\kern0.4\wd0\vrule 
height0.7\ht0\hss}\box0}}
{\setbox0=\hbox{$\scriptscriptstyle\rm Q$}\hbox{\raise 0.15\ht0\hbox 
to0pt{\kern0.4\wd0\vrule
height0.7\ht0\hss}\box0}}}}
\def\nbT{{\mathchoice {\setbox0=\hbox{$\displaystyle\rm 
T$}\hbox{\hbox to0pt{\kern0.3\wd0\vrule
height0.9\ht0\hss}\box0}} {\setbox0=\hbox{$\textstyle\rm 
T$}\hbox{\hbox to0pt{\kern0.3\wd0\vrule
height0.9\ht0\hss}\box0}} {\setbox0=\hbox{$\scriptstyle\rm 
T$}\hbox{\hbox to0pt{\kern0.3\wd0\vrule
height0.9\ht0\hss}\box0}} {\setbox0=\hbox{$\scriptscriptstyle\rm T$}\hbox{\hbox
to0pt{\kern0.3\wd0\vrule height0.9\ht0\hss}\box0}}}}
\def\nbS{{\mathchoice {\setbox0=\hbox{$\displaystyle     \rm 
S$}\hbox{\raise0.5\ht0%
\hbox to0pt{\kern0.35\wd0\vrule height0.45\ht0\hss}\hbox 
to0pt{\kern0.55\wd0\vrule
height0.5\ht0\hss}\box0}} {\setbox0=\hbox{$\textstyle        \rm 
S$}\hbox{\raise0.5\ht0%
\hbox to0pt{\kern0.35\wd0\vrule height0.45\ht0\hss}\hbox 
to0pt{\kern0.55\wd0\vrule
height0.5\ht0\hss}\box0}} {\setbox0=\hbox{$\scriptstyle      \rm 
S$}\hbox{\raise0.5\ht0%
\hboxto0pt{\kern0.35\wd0\vrule height0.45\ht0\hss}\raise0.05\ht0%
\hbox to0pt{\kern0.5\wd0\vrule height0.45\ht0\hss}\box0}} 
{\setbox0=\hbox{$\scriptscriptstyle\rm
S$}\hbox{\raise0.5\ht0%
\hboxto0pt{\kern0.4\wd0\vrule height0.45\ht0\hss}\raise0.05\ht0%
\hbox to0pt{\kern0.55\wd0\vrule height0.45\ht0\hss}\box0}}}}
\def\nbZ{{\mathchoice {\hbox{$\sf\textstyle Z\kern-0.4em Z$}} 
{\hbox{$\sf\textstyle Z\kern-0.4em Z$}}
{\hbox{$\sf\scriptstyle Z\kern-0.3em Z$}} 
{\hbox{$\sf\scriptscriptstyle Z\kern-0.2em Z$}}}}

\begin{document}

\title{ {Functional  renormalization group approach to non-collinear  magnets}}

\author{B. Delamotte} \email{delamotte@lptmc.jussieu.fr}
\affiliation{Sorbonne Universit\'es, UPMC Univ  Paris 06,  LPTMC,  CNRS  UMR 7600, F-75005, Paris, France}

\author{M. Dudka} \email{maxdudka@icmp.lviv.ua} 
\affiliation{Institute  for Condensed   Matter  Physics,  National  Acad.  Sci.   of
Ukraine, UA--79011 Lviv, Ukraine}

\author{D. Mouhanna} \email{mouhanna@lptmc.jussieu.fr}
\affiliation{Sorbonne Universit\'es, UPMC Univ  Paris 06,  LPTMC,  CNRS  UMR 7600, F-75005, Paris, France}

\author{S. Yabunaka}\email{yabunaka@scphys.kyoto-u.ac.jp}
\affiliation{Yukawa Institute for Theoretical Physics, Kyoto University, Kyoto 606-8502, Japan}

\begin{abstract}

A functional renormalization group approach to  $d$-dimensional,  $N$-component,  non-collinear  magnets 
is performed using various  truncations  of the effective action  relevant  to study their 
long distance behavior. With help of these truncations we  study  the existence  of a  stable 
fixed point  for dimensions  between  $d= 2.8$ and $d=4$  for various values of   
$N$ focusing on the  critical value, $N_c(d)$,  that, for a given dimension $d$, 
separates  a first order region  for $N<N_c(d)$ from a second order region for $N>N_c(d)$. 
Our approach concludes  to  the {\it absence}  of   stable fixed point  in the physical 
-- $N=2,3$ and $d=3$ -- cases, in agreement  with $\epsilon=4-d$-expansion and  in contradiction with previous  
perturbative approaches  performed at  fixed dimension  and  with recent approaches   based on conformal  bootstrap program. 

\end{abstract}

\vskip 1cm

\maketitle

\section{Introduction}

Non-collinear -- or canted  -- XY or Heisenberg magnets are one of the simplest examples  
of physical systems for which  the order parameter  is not a {\it vector},  as in the collinear 
case, but a  $2nd$-order tensor, that is, a  {\it matrix} \cite{delamotte03}. As a   
consequence, the rotation group  is fully broken in the  low temperature phase. 
This is particularly clear for the stacked triangular  antiferromagnetic (STA)  spin system,  since, 
in the ground state, the spins on a plaquette  exhibit  
the famous  120$^{\circ}$ structure  which {\it  completely}  breaks the $SO(N)$ rotation group with  $N=2$ or 3.
This change of symmetry-breaking pattern between  collinear and non-collinear
magnets alters   drastically  the critical physics which is not yet fully clarified  despite forty years  
of intensive research (see \cite{delamotte03} and references therein).

On the experimental side   there is  agreement   that at the phase transition, both XY and Heisenberg non-collinear magnets 
exhibit  scaling laws but with critical exponents differing   from those of the $O(N)$  model. This  has lead to two kinds of scenarios: 
either the phase transitions belong to a new universality class \cite{kawamura88} or they are {\it weakly}  of first order  \cite{delamotte03}.
A  careful analysis of  the results coming from different  materials shows  that  the critical exponents 
vary from one compound to the other and there is in fact no universality  \cite{delamotte03}. This tends to confirm the assumption 
of weak first order  transitions as do  almost  all  numerical  simulations  performed  on STA or on similar  
models  \cite{diep89,loison98,loison00b, itakura03,peles04,bekhechi06,quirion06,zelli07,thanhngo08}
but one  \cite{calabrese04}. However all these results do not rule out definitively the possibility that some 
other systems undergo a second order phase transition.

On the theoretical side, all approaches agree that in the vicinity of four dimensions, there exists a line  $N_c(d)$
in the $(d,N)$ plane above which the transition is of second order and below which it is of first order.
At two loops,   $N_c(d=4-\epsilon)=21.8-23.4 \epsilon + O(\epsilon^2)$ \cite{jones76,bailin77}, 
which shows that perturbative corrections are  large  and that  it is therefore impossible to extrapolate 
reliably this result in $d=3$ without computing the contributions of higher orders. The perturbative calculation of $N_c(d)$ 
either within the $\epsilon$ or pseudo-$\epsilon$-expansion \cite{jones76,bailin77,garel76,barak82,antonenko95,calabrese03c,holovatch04} 
or directly in $d=3$ \cite{calabrese04,antonenko94,pelissetto01a,calabrese02,calabrese03b}
has been performed up to five or  six loops. The same quantity has also been computed within the
nonperturbative renormalization group (NPRG) approach \cite{zumbach93,zumbach94,zumbach94c,tissier00,tissier00b,tissier01, tissier03, delamotte03}.

However  the various results so far obtained do not  agree in particular in the physically interesting   $N=2$ and $N=3$ cases.
On the one hand, the $\epsilon$ and pseudo-$\epsilon$-expansion and the NPRG approach  both lead to a value of $N_c(d=3)$ well above 3. Indeed:  $N_c(d=3)=6.1(6)$ within the $\epsilon$-expansion at five loops \cite{calabrese03c}, $N_c(d = 3)=6.22(12)$   \cite{calabrese03c} and 
$N_c(d = 3)=6.23(21)$ \cite{holovatch04}  within the pseudo-$\epsilon$-expansion at six loops and finally   $N_c(d=3)=5.1$  within the NPRG approach \cite{delamotte03}.  Moreover the  NPRG approach  predicts that the transitions are weakly of first
order for $N=2$ and 3,  accompanied by a non-universal scaling behavior in agreement with both numerical computations and experiments. 
On the other hand,  perturbative computations  performed at fixed dimension lead to completely different 
predictions \cite{calabrese04,calabrese02,calabrese03b}.   Those   performed at six loops  
within the zero momentum massive scheme \cite{calabrese03b} lead  to two different $N_c(d=3)$: $N_{c1}$ and $N_{c2}$.
Above  $N_{c1}=6.4(4)$ there are, as usual, two fixed points, one stable and one unstable. They get closer when $N$ is lowered,   collide at $N_{c1}$ and disappear. However, below $N_{c2}=5.7(3)$, a stable fixed point reappears  and  exists for $N=2$ and 3. A second order phase transition is
therefore predicted  for these values of $N$. A puzzling  feature of this  fixed-dimension computation  is that  it  predicts, in the $(d,N)$ plane,  
a curve $N_c(d)$ showing a very unusual S-like shape around $d=3$ or, equivalently, a {\it non-monotonic}  curve $d_c(N)$,   the dimension below which, for a  given $N$ the transition is of second order. If true, this would imply that at $d=3$, a second order
transition occurs for low values of $N$ whereas at larger values it would be of first order before being  {\it again} of second order.  As for  computations performed at five loops  within  the  minimal subtraction ($\overline{MS}$) scheme  {\it without}  $\epsilon$-expansion  \cite{calabrese04},  
  no critical value of  $N$ is found in $d=3$ and,  thus, a  second order phase transition is predicted for {\it any}
value of $N$.

In order to solve the contradiction between the different approaches, a careful re-analysis of the resummation 
procedures used within  fixed-dimension calculations has  been performed in  \cite{delamotte08,delamotte10,delamotte10b}. 
Some  very peculiar features  have been revealed by   this analysis: lack of convergence of the 
critical quantities with the order of the expansion,   high sensitivity  of the results with respect  to the 
resummation parameters,  existence  of the non-trivial fixed point {\it at}Ê and  {\it above}  the 
upper  critical dimension $d=4$, etc.  This re-analysis have cast grave doubts upon  the  
reliability  of  the  fixed-dimension approaches  when applied to non-collinear  magnets.

From the results above, it is tempting  to believe that  a   coherent picture of the critical behavior 
of non-collinear magnets has been   reached and that the transitions are of (weak) first order  whereas  fixed-dimensions approaches would be not converged or  deficient.  However,  recently,  a new  approach,  based on the  conformal 
bootstrap (CB) program (see for instance \cite{kos15} and references therein), 
has been  applied to  $O(N)\times O(M)$ models \cite{nakayama14a,nakayama14b}, the  $M=2$ case corresponding  
to non-collinear magnets. This program  when applied to  other systems such as the three-dimensional 
ferromagnetic Ising model \cite{elshowk12a,elshowk14b}  leads to the best determination of the critical 
exponents ever obtained. It also has the advantage of being {\it a priori} unbiased by convergence problems 
since it is not based on series expansions, contrary to RG methods. As for the $O(N)\times O(M)$  models,
the authors of  \cite{nakayama14b} have found  with the CB approach a stable fixed point  
for  $N=3$ and  $M=2$ in $d=3$   with critical exponents in  good agreement with the fixed-dimension perturbative 
results  and thus in disagreement with both the  $\epsilon$, pseudo-$\epsilon$-expansion and NPRG results. However, it is important to notice two points.   First, the CB program assumes scale invariance,  that is, the existence of a second order phase transition 
which is  precisely the main  question addressed in the context of  non-collinear  magnets.   
Second, in all  RG-based schemes, there   exists, in the RG flow diagram,  either two  nontrivial fixed points or none. Indeed   the very mechanism that determines the curve  $N_c(d)$   when $N$ is lowered keeping $d$ fixed is the collapse 
of the stable -- chiral --   fixed point $C_+$ controlling the phase transition with another 
unstable -- anti-chiral -- fixed  point   $C_-$.  However,  this fixed point  $C_-$ is  not  found    
within  the CB analysis of  the $O(3)\times O(2)$ model \cite{nakayama14b},  
which  contrasts  with  the other approaches. 

The   results obtained within  the CB  program  have led  us to re-examine the NPRG approach to 
non-collinear magnets, looking for possible failures in  previous computations \cite{zumbach93,zumbach94,zumbach94c,tissier00,tissier00b,tissier01, tissier03, delamotte03}  
that  were based on two kinds of approximation:  (i)  the derivative expansion of the
action  at  its lowest order, called the local potential approximation  (LPA)  \cite{zumbach93,zumbach94,zumbach94c};   (ii) a field 
expansion of   the effective action   including the  effects of derivative terms at leading order \cite{tissier00,tissier00b,tissier01, tissier03, delamotte03}. 

In this article, we go beyond   these  approximations     by performing:   (i)  an approximation -- called  field-semi-expansion  --  where  
the most  important   field-dependence is considered  without approximation, {\it i.e.} functionally, the remainder  being approximated 
using a field-expansion;    (ii)  an  approximation where  no  field-expansion at all is performed. Moreover  in these 
two approximations  the most important  derivative terms are taken into account at leading order. The results of our analysis  
corroborate all  conclusions reached  within  previous NPRG approaches in particular for the value of 
$N_c(d=3)$ that we find to be significantly  larger than 3. They  thus 
disagree with those obtained with the fixed-dimension  perturbative  approaches  as well as  with  the CB program. 

The article is organized as follow.  In  section  II, we briefly recall the principle and properties   of the NPRG approach that 
we use.  In section III,  we present  the effective action relevant for non-collinear magnets within this approach.  
We discuss in particular  its  symmetries,  the symmetry breaking scheme and, finally, the approximations  used. In section  IV,
we  present  the  NPRG equations for  
non-collinear magnets.  In section V, we give  the results obtained  under the form a field-semi-expansion of the associated  
effective potential. In section VI, we discuss the results obtained using a full functional  approximation of the 
effective action where only the derivative terms are truncated.  In section VII, we conclude. 

\section{The effective action method}

The central object  of  the  NPRG  approach is a   running  
effective  action  -- or Gibbs free energy  -- $\Gamma_k$  
 \cite{wetterich91,wetterich93b}    that includes the 
statistical fluctuations  between the typical  microscopic momentum scale $\Lambda$ 
of the system -- the inverse of a lattice spacing for instance -- down to the  running scale $k<\Lambda$.
In the limit   $k\to\Lambda$,  no fluctuation is taken into account  and $\Gamma_{k=\Lambda}$ identifies  with  
the  classical action  -- or  microscopic Hamiltonian --  while when $k\to 0$, all fluctuations are summed over 
and the usual  Gibbs free energy $\Gamma$ is recovered:  
\begin{equation}
\left\{
\begin{array}{l}
\Gamma_{k=\Lambda}=S\ \\ 
\\
\Gamma_{k=0}=\Gamma\  .
\label{limitesgamma}
\end{array}
\right.
\end{equation}
Thus,  at any  finite scale $k<\Lambda$,  $\Gamma_k$  interpolates between 
the action  and the Gibbs free energy.  To construct the  running effective action, the  original partition function : 
\begin{equation}
 {\cal Z}[J]=\int  D\zeta \exp\Big(-S[\zeta]+ \int_q J(q)\zeta(-q)\Big)
\end{equation}
where $\int_q=\int d^dq/(2\pi)^d$, is  modified  by adding a  cut-off  term to  the   classical action: 
\begin{equation}
\Delta S_k[\zeta]={1\over 2} \int_ q \
{R}_k(q^2)\  \zeta(q) \zeta(- q)
\label{massterm}
\end{equation}
in which $R_k(q^2)$ is  a  cut-off function that ensures the separation between the low- and high-momentum  modes.
The $k$-dependent partition function thus writes:
\begin{equation}
 {\cal Z}_k[J]=\hspace{-0.1cm}\int D\zeta \exp\Big(-S[\zeta]-\Delta S_k[\zeta]+ \int_q J(q)\zeta(-q)\Big) \  .
\end{equation}

It is convenient to choose $R_k(q^2)$ such that:  (i)  it behaves as a  mass  at low momentum  $q$ 
in order to freeze the low-momentum fluctuations;  (ii) it vanishes at  large  momentum $q$ in order to keep unchanged the high-momentum 
fluctuations.  Thus one has:
\begin{equation}
\left\{
\begin{array}{l}R_k(q^2)\sim k^2 \hspace{0.5cm} \hbox{when} \hspace{0.5cm}  q^2\ll k^2
\\
\\
R_k(q^2)\to 0 \hspace{0.6cm} \hbox{when}
\hspace{0.5cm} {q}^2\gg k^2 \ .
\end{array}
\right.
\end{equation}
These constraints on $R_k(q^2)$ also  imply that $R_{k=0}(q^2)\equiv 0$ which is consistent with
the fact that when $k=0$ all fluctuations have been summed over and the original model is retrieved: ${\cal Z}_{k=0}[J]={\cal Z}[J]$.
Finally note that  $R_{k=\Lambda}(q^2)$ is very large for  all momenta much smaller than $\Lambda$. 
A typical cut-off function  satisfying all the previous requirements is:
\begin{equation}
R_k(q^2)={Z_k {q}^2\over e^{{q}^2/k^2}-1}\ 
\label{cutoffexp}
\end{equation}
where $Z_k$ is the field renormalization -- see below.  Another useful cut-off function,  called theta cut-off,  
has been   proposed by Litim \cite{litim02}. It  writes:
\begin{equation}
R_k(q^2)=Z_k  \left(k^2- {q}^2\right)\,\Theta\left(k^2-{q}^2\right)
\label{cutoffstep}
\end{equation}
where $\Theta$ is the usual step function.
 
The  running Gibbs free energy $\Gamma_k$  is then defined as the (slightly modified)  Legendre transform of the Helmoltz  
free  energy $W_k[J]=\log{\cal Z}_k[J]$ 
(see \cite{delamotte03,berges02,pawlowski07,kopietz10,rosten12,nagy14}):
\begin{equation}
\Gamma_k[\phi]=-W_k[J]+ J.\phi -\Delta S_k[\phi]
\label{defgamma}
\end{equation} 
where  $\phi$ is the order parameter field and where a mass term analogous to (\ref{massterm})  has 
been added  with respect  to the usual definition of $\Gamma$ for the following reason.  Clearly,  since $R_{k=0}(q^2)\equiv 0$,  
with the  definition Eq.(\ref{defgamma})  one recovers  the  usual free  energy in the limit  $k\to 0$: 
$\Gamma_{k\to 0}=\Gamma$.  The  limit   $k\to \Lambda$ is less trivial since, there, the cut-off function 
$R_k(q^2)$ is very large.  But one can show \cite{delamotte03,berges02,pawlowski07,kopietz10,rosten12,nagy14} 
 that the  cut-off term $\Delta S_k[\phi]$  in Eq.(\ref{defgamma})  conspires  with that included in $W_k[J]$ 
and makes  that  $\Gamma_{k=\Lambda}[\phi]\simeq S[\phi]$. One thus recovers  the conditions Eqs.(\ref{limitesgamma}).

 The effective action  follows  an exact  flow equation,  the Wetterich equation \cite{wetterich93c,ellwanger93b,tetradis94,morris94a}:
\begin{equation}
\partial_t \Gamma_k{[\phi]}= \displaystyle {1\over 2}{\rm Tr}\int_{q} \ \displaystyle
{\dot R_k(q^2) \left(\Gamma_{k}^{(2)}[q,-q, \phi]+R_k(q^2)\right)^{-1}}
\label{Wetterichfinal}
\end{equation}
where  $t=\ln {k/ \Lambda}$ and $\dot{R_k}=\partial_t R_k$.  In Eq.(\ref{Wetterichfinal}), $\hbox{Tr}$ 
must be understood  as a trace over  internal vector or tensor indices  if  the order parameter  $\phi$ spans 
a nontrivial representation of a group which is the 
case for non-collinear magnets. Finally $\Gamma_{k}^{(2)}[q,- q,\phi]$ stands for the Fourier transform of  the second functional derivative  of 
$\Gamma_k$ with respect to the order parameter field:
\begin{equation}
 \Gamma_{k,i,j}^{(2)}[x,y,\phi]=\frac{\delta \Gamma_k[\phi]}{\delta\phi_i(x)\delta\phi_j(y)}
\end{equation}      
for a $N$-component  field  with  components  $\phi_i$.  Thus  the quantity $(\Gamma_k^{(2)}[\phi]+ R_k)^{-1}$ 
appearing in Eq.(\ref{Wetterichfinal}) represents the {\it exact}, {\it i.e.} field-dependent,  propagator.   

The general properties of  Eq.(\ref{Wetterichfinal})   have been discussed  at length in the literature 
 \cite{delamotte03,berges02,pawlowski07,kopietz10,rosten12,nagy14}.   We recall only some of them that are 
directly relevant for our purpose.   First, Eq.(\ref{Wetterichfinal}) is exact,  notably  because the propagator $(\Gamma_k^{(2)}[\phi]+ R_k)^{-1}$ is the exact, field-dependent one.   As a consequence, Eq.(\ref{Wetterichfinal})  embodies  all perturbative and nonperturbative  features of 
the model under study:  spin-waves, topological excitations, bound states, tunneling, etc.  Second,  due  to the 
property of the cut-off function $R_k$,  Eq.(\ref{Wetterichfinal})  is,   due to the  presence of a ``mass-term"  
$R_k$ in the propagator,  infrared  finite   for any $k>0$.  Also, due to  presence of  the function $\dot R_k$  
that  rapidly decays for  high momenta, {\it i.e.} ${q}^2>k^2$,  it is ultraviolet finite. Thus  Eq.(\ref{Wetterichfinal})  
allows  to explore  criticality  directly in three dimensions without having recourse to tricks like  $\epsilon$-expansion 
techniques for instance. Third, Eq.(\ref{Wetterichfinal}) has a one-loop structure which  implies that all integrals 
encountered are single integrals  in  contrast  with perturbative expansions  at high orders that lead to involved 
multiple integrals. This  property makes straightforward  comparisons  with leading order of  {\it all}
perturbative approaches : weak coupling, low temperature, large-$N$,  expansions in their respective domains  of validity. 
 
 Now, although exact,  Eq.(\ref{Wetterichfinal})  must be approximated in order to get concrete results for complicated problems with optimal efforts. This is 
 realized by choosing a  truncation  for $\Gamma_k{[\phi]}$.  Among the most popular ones one finds,  for a  scalar field theory: 
 
 (i)  {\sl Derivative expansion.}  $\Gamma_k$ is expanded in powers of the derivatives of the order parameter:  
\begin{equation}
 \displaystyle\Gamma_k[\phi]=\int_x \;\bigg\{{U_k(\phi)}+{1\over 2}\ {Z_k(\phi)}\:
(\partial \phi)^2 + {\cal{O}}((\partial \phi)^4) \bigg\}\ 
\end{equation}
from which one gets  {\it functional} RG equations for the potential part $U_k(\phi)$ and  
kinetic part  $Z_k(\phi)$ of the running effective action by appropriate functional derivations 
of Eq.(\ref{Wetterichfinal}). The rationale behind this  approximation  is that when  the  anomalous 
dimension is   small,   gradient terms  should play only a small role at 
long distance and high-order  derivative terms should be negligible. 

\vspace{0.2cm} 
  
(ii)  {\sl  Combined derivative and field expansions.}   On top of the derivative expansion, the functions $U_k(\phi)$ and $Z_k(\phi)$ are 
 expanded  in powers of $\phi$ around a given field  configuration. This  kind of approximation  converts  the functional equation Eq.(\ref{Wetterichfinal})  
 into a set of ordinary coupled differential equations for the coefficients of the expansion. This approximation  
 relies on    the double assumption that the anomalous dimension is small and that the 
 correlation functions with a high number of legs have a small back-reaction in the RG flow on those  
 with a small number of legs.  Neglecting them should therefore not spoil the dominant critical behavior.
 
 \vspace{0.2cm}

(iii)  {\sl Field expansion.} $\Gamma_k$ is expanded in powers of the order parameter $\phi$. One has:
\begin{equation}
\hspace{-0.2cm} \Gamma_k[\phi]=\sum_{n=0}^{\infty} {1\over n!}  
\left( \int_{x_i} \prod_{i=0}^{n}\phi({x}_i)\right) {\cal V}_k^{(n)}({ x}_1,...,{x}_n)
\end{equation}
where ${\cal V}_k^{(n)}({x}_1,...,{x}_n)$  are  the vertices of the theory. Applying then 
Eq.(\ref{Wetterichfinal}) to this expansion allows to generate a hierarchy of RG equations for the vertices. 
Approximations  are  then performed on $\Gamma_k$ in order  to close  this hierarchy. 
This kind of approach relies on  the assumption  that  vertices of high orders  
in the field  can be neglected while the full  functional dependence with respect to derivatives -- or momenta -- 
must be kept.  This is very likely not the generic situation.  

\vspace{0cm} 

 (iv)  {\sl Green function}, also called   {\sl Blaizot--M\'endez-Galain--Wschebor (BMW)}  \cite{blaizot06,blaizot06b,blaizot06c},  approach.  
 Its aim is to keep the full momentum dependence of the two-point functions and the functional dependence of the 
 potential $U_k(\phi)$ and $Z_k(\phi)$ functions. It consists in approximating the momentum dependence of the three- and four-point functions
 in the flow of the two-point functions. It thus reproduces the results obtained within the derivative expansion
 at small momenta while being also non-perturbative at high momenta.

From a practical point of view the main  difficulty   lies in the choice  of  approximation  that  contains the most important  features 
of the model under study while being  also technically  manageable.  For critical phenomena  one can focus on momenta $q$ close 
to 0.   The method of choice is therefore  the derivative expansion (see (i) above).   
This is in contrast with  situations where  bound states  -- or, more generally,   excitations exhibiting  a non-trivial 
momentum structure  -- are expected in which  a more or less important momentum  
dependence should be considered through the use of (iii) or  (iv).    At  leading order  of the derivative expansion  
one completely neglects the  effects of derivative terms and sets the  field renormalization function  $Z_k(\phi)$ 
equal to one, keeping a full function $U_k(\phi)$ for the potential part.  This is the so-called 
local potential approximation (LPA).  A possible improvement of  this approach consists  in  replacing now $Z_k(\phi)$ 
by a non-trivial constant   $Z_k$   from which follows  a   $k$-dependent anomalous dimension $\eta_k=-(1/Z_k) \partial_t Z_k$, 
the usual anomalous dimension being given by $\eta_k$ at a  fixed point. This  approximation  is sometimes called the LPA'. 
This is essentially this approach that  is employed in this article.  Note for completeness  that one can also  try to 
treat  the kinetic terms  $Z_k(\phi)$  as a function. However,  in practice this has only been performed in the simplest 
case of $O(N)$ models \cite{morris98d}.  For more involved models  only approximation (ii) has been used. 
This approach, when used at reasonably  high orders in the field-expansion,  is in general sufficient 
to obtain high  precision results.  For  the Ising model for instance,  the best estimates of the critical exponents in $d=3$ have been recovered with
this combined  derivative and field  expansion employed at fourth order in the derivative and tenth order  in the field  \cite{canet03b}.

 Finally, let us   emphasize that while  the accuracy of the results obviously depend on  the order of the truncations 
 none  of the approximations  presented above spoils the nonperturbative character of the method.  
 Indeed, although the effective action itself  is approximated, the very structure of Eq.(\ref{Wetterichfinal}) 
 is kept unchanged  as far as the  left-hand side is not  expanded  in powers of one of the   usual perturbative parameters: 
 coupling constant, temperature, or $1/N$. Thus even at the lowest order of the combined derivative and field-expansion one 
 already gets results unreachable by perturbative methods \cite{delamotte03,berges02,pawlowski07,kopietz10,rosten12,nagy14}.

\section{The effective action for non-collinear  magnets}

We now present  the derivative expansion of  the effective action relevant to  non-collinear magnets. 
First we recall that  for $N$-component non-collinear magnets,   the order parameter consists of  
two $N$-component real vectors  $\vec \phi_1$ and $\vec \phi_2$ which, in the STA model, represent  
linear combinations of the spins of a plaquette \cite{delamotte03}.   It is convenient to  gather 
these two fields  into  a   $N\times 2$ matrix:   $\Phi=(\vec\phi_1,\vec\phi_2)$.   In the  continuum  
limit, the action for non-collinear magnets displays a $O(N)\times O(2)$  invariance  where $O(N)$ stands 
for the usual  rotational invariance  while $O(2)$ reflects the original $C_{3v}$ symmetry of the lattice \cite{delamotte03}.    
The left $O(N)$ and right $O(2)$ transformations are implemented on $\Phi$ by:
\begin{equation}
\left\{
\begin{array}{ll}
&\Phi \to U\Phi, \ U\in O(N)\\
\\
&\Phi\to \Phi V, \ V\in O(2)\ .
\end{array}
\right.
\end{equation}
As said above  the derivative expansion consists in expanding  $\Gamma_k$ in powers of $\partial\phi$  at a finite order.  We  consider here  the expansion at
second order in derivatives where  $\Gamma_k$  writes\cite{delamotte03}: 
\begin{widetext}
\begin{equation}
\begin{split}
&\Gamma_k[\vec \phi_1,\vec \phi_2]=\int_x \bigg\{ U_k(\rho,\tau)+\frac{1}{2}
Z_k(\rho,\tau)\Big(\big(\partial \vec \phi_1\big)^2+ \big(\partial \vec
\phi_2\big)^2\Big) + \frac{1}{4} Y^{(1)}_k(\rho,\tau) \big(
\vec\phi_1\cdot\partial \vec\phi_2- \vec\phi_2\cdot\partial
\vec\phi_1\big)^2+ \\ & +\frac 14 Y^{(2)}_k(\rho,\tau) \big(
\vec\phi_1\cdot\partial \vec\phi_1+ \vec\phi_2\cdot\partial
\vec\phi_2\big)^2 + \frac 14 Y^{(3)}_k(\rho,\tau) \Big(
\big(\vec\phi_1\cdot\partial \vec\phi_1- \vec\phi_2\cdot\partial
\vec\phi_2\big)^2+ \big(\vec\phi_1\cdot\partial \vec\phi_2+
\vec\phi_2\cdot\partial \vec\phi_1\big)^2\Big) \bigg\} \ .
\label{actiongen}
\end{split}
\end{equation}
\end{widetext}
In Eq.(\ref{actiongen}),  $\rho$ and $\tau$ are the  two independent $O(N)\times O(2)$ invariants  that   
read in terms of $\vec \phi_1$ and $\vec \phi_2$:
\begin{equation}
\left\{
\begin{array}{ll}
\rho&=\hbox{Tr}(^{t}\Phi.\Phi)={\vec \phi_1}^{\, 2}+{\vec \phi_2}^{\, 2}\\
\\
\tau\hspace{-0mm}&=\displaystyle\frac{1}{2}\hbox{Tr} \left(^t\Phi.\Phi-\openone\ \displaystyle {\rho\over 2}\right)^2\\
\\
&=\displaystyle{1\over 4} \Big({\vec \phi_1}^{\, 2}-{\vec \phi_2}^{\, 2}\Big)^2+({\vec \phi}_1\cdot{\vec \phi}_2)^{\, 2}\ .
\end{array}
\right.
\end{equation}

The term $U_k(\rho,\tau)$ in Eq.(\ref{actiongen})   
represents  the potential part of the effective action  while $Z_k(\rho,\tau)$ and $Y_k^{(i)}(\rho,\tau)$, $i=1,2,3$, 
are kinetic parts.  At the minimum of the potential $U_k(\rho,\tau)$   the vectors $\vec\phi_1$ and  
$\vec\phi_2$   are orthogonal with the same norm, which corresponds to the following configuration:
\begin{equation}
\Phi_0=
\begin{pmatrix}
\phi&0\\
0&\phi\\
0&0\\
\vdots&\vdots\\
0&0
\end{pmatrix}
\label{groundstate}
\end{equation}  
where $\phi$ is a constant.
For the spins on the lattice, the  configuration Eq.(\ref{groundstate})  corresponds to the 120$^{\circ}$ structure. 
Notice that $\tau$ has been built such that it vanishes in this configuration. The ground state Eq.(\ref{groundstate})   
is invariant  under the   $O(N-2)$  group of 
left rotations and a diagonal $O(2)$ group --  $O(2)_{\hbox{\scriptsize diag}}$ -- that combines left and right rotations: 
\begin{equation}
\left(
    \begin{array}{cc}
 O(2)
    &0\\
    0&O(N-2)
    \end{array}\right)
   \Phi_0\    O(2)    = \Phi_0 \ . 
      \end{equation}
The symmetry breaking scheme 
is thus given by: $O(N)\times O(2)\to O(N-2)\times O(2)_{\hbox{\scriptsize diag}}$.  For $N=3$   one  recovers  the symmetry breaking scheme:
\begin{equation}
 G=O(3)\times O(2) \to H={\nbZ}_{2}\times O(2)_{\hbox{\scriptsize diag}}\ 
\end{equation}
which shows that the  order  parameter space is   given by  $SO(3)$.  For $N=2$ the symmetry breaking scheme is given by:
\begin{equation}
G=O(2)\times O(2) \to H=O(2)_{\hbox{\scriptsize diag}}
\end{equation}
or simply by $SO(2)\times \nbZ_2\to \openone$  where the  degrees of freedom associated with the $ \nbZ_2$  group are  
referred to as chirality  excitations.  

We now proceed to further approximations. As explained in \cite{delamotte03},   only the  functions  $Z_k$ and 
$Y_k^{(1)}$  contribute to  the physics of Goldstone modes and thus are supposed to be the  most  relevant at the transition.  
Thus  we  consider the following simplified effective action:
\begin{eqnarray}
\label{action_generale2}
\Gamma_k&&\hspace{-0cm}=\int_x \Big\{U_k(\rho,\tau)+\frac{1}{2}
Z_k(\rho,\tau)\Big(\big(\partial \vec \phi_1\big)^2+ \big(\partial \vec
\phi_2\big)^2\Big)\nonumber\\
&&+ \frac{1}{4} Y^{(1)}_k(\rho,\tau) \big(\vec\phi_1\cdot\partial \vec\phi_2- \vec\phi_2\cdot\partial
\vec\phi_1\big)^2 \Big\} \ .
\end{eqnarray}
A last  approximation consists   in neglecting the field-dependence of the  functions $Z_k(\rho,\tau)$ and $Y^{(1)}_k(\rho,\tau)$
and, thus, in setting $Z_k(\rho,\tau)=Z_k$ and $Y^{(1)}_k(\rho,\tau)=\omega_k$. 
This approximation  has been used in \cite{tissier00,tissier00b,tissier01, tissier03, delamotte03}  
in which  the function $U_k(\rho,\tau)$ was further expanded in powers of the invariants $\rho$ and $\tau$. 
All perturbative  results were recovered this way, that is, the one-loop results obtained  either in 
$d=4-\epsilon$ or  in $d=2+\epsilon$. Note that around $d=2$, it is  necessary to take into account the so-called ``current-term"  
-- $(\vec\phi_1\cdot\partial \vec\phi_2- \vec\phi_2\cdot\partial \vec\phi_1)^2$ 
--  to get the phenomenon of enlarged  symmetry at the fixed point \cite{azaria90}, although it is irrelevant 
by power-counting in the vicinity of $d=4$. In  $d=3$, the dimension in which we are interested in this article, such a term 
should not  contribute significantly. We however keep it as we are tracking possible weaknesses of previous NPRG  approaches.

\section{Renormalization Group equations}

We now  present the RG equations relevant  to  non-collinear magnets.  In the case of a model 
with $O(N)\times O(2)$ symmetry  the exact flow equation Eq.(\ref{Wetterichfinal}) writes:
\begin{equation}
\partial_t \Gamma_k{[\vec \phi_1,\vec \phi_2]}= \displaystyle {1\over 2}{\rm Tr}\int_{q} \ \displaystyle
{\dot R_k({q}) \left(\Gamma_{k}^{(2)}[{q},-{q},\vec \phi_1,\vec \phi_2]+R_k\right)^{-1}}
\label{eqrg}
\end{equation}
where $\Gamma_{k}^{(2)}[{q},-{q},\vec \phi_1,\vec \phi_2]$ is the Fourier transform of the second functional derivatives of $\Gamma_{k}$:
\begin{equation}
 \Gamma_{k,(a,i),(b,j)}^{(2)}[x,y,\vec \phi_1,\vec \phi_2]=\frac{\delta \Gamma_k[\vec \phi_1,\vec \phi_2]}{\delta\phi_a^i(x)\delta\phi_b^j(y)}
 \label{secondderivatives}
\end{equation}
where $a,b=1, 2$ and $i,j=1,\cdots,N$.

The flow equation for the effective potential  $U_k(\rho,\tau)$ follows from its definition:
 \be
 {U}_k(\rho,\tau)= {1\over \Omega}\Gamma_k{[\vec \phi_1,\vec \phi_2]}\Big\vert_{\Phi}
  \ee
where $\Omega$ is the volume of the system and $\Phi$ a constant   field configuration.  
Since $U_k$ is an O$(N)\times $O(2) invariant, it is possible to use O$(N)\times $O(2) transformations to simplify as much as possible
the configuration $\Phi$ in which its  RG  flow   (\ref{eqrg}) is evaluated. It  is easy to show  using  
these transformations that one can  to recast any constant matrix $\Phi$ in
a diagonal ``form":
\begin{equation}
\Phi=U \Phi_D V \hspace{0.2cm} \hbox{with} \hspace{0.2cm} U\in O(N)  \hspace{0.2cm}  \hbox{and}  \hspace{0.2cm}  V\in O(2)
\end{equation}
with:
\vspace{-0.2cm}
\begin{equation}
\Phi_D=
\begin{pmatrix}
\phi_1&0\\
0&\phi_2\\
0&0\\
\vdots&\vdots\\
0&0
\end{pmatrix}
\label{confgeneric}
\end{equation} 
where $\phi_1$ and $\phi_2$ are constants. 

The -- derivative --  coefficients  $Z_k$ and $\omega_k$ are defined by: 
\begin{equation}
Z_k=\frac{(2\pi)^d}{\delta(0)}\lim_{p^2\to
0}\frac{d}{dp^2}\left(\left.\frac{\partial^2 \Gamma_k}{\delta
\phi_1^1({p})\delta \phi_1^1(-{p})}\right|_{\Phi_I}\right)
\label{defZ}
\end{equation}
\begin{equation}
\frac{\omega_k}{2}=\frac{(2\pi)^d}{{\kappa}   \delta(0)}\lim_{p^2\to
0}\frac{d}{dp^2}\left(\left.\frac{\partial^2 \Gamma_k}{\delta
\phi_1^2({p})\delta \phi_1^2(-{p})}\right\vert_{\Phi_I}\right)-\frac{Z_k}{{\kappa}},
\label{defomega}
\end{equation}
where we choose   a  uniform  field configuration $\Phi_I$  proportional to the identity:
\begin{equation}
\Phi_I=
\begin{pmatrix}
\sqrt{\kappa}&0\\
0&\sqrt{\kappa}\\
0&0\\
\vdots&\vdots\\
0&0
\end{pmatrix}
\label{diagconf}
\end{equation}   
$\sqrt{\kappa}$ being  a constant that can be different from  the minimum $\phi$ of the potential, Eq.(\ref{groundstate}). 
Finally the running anomalous   dimension  $\eta_k$ is  defined by   $\eta_k=-\partial_t \log Z_k$. 
  
The flow equation for the potential writes in terms of the various propagators associated 
with the mass spectrum of the model  (see Appendix (\ref{propagator})):
\begin{widetext}
\begin{eqnarray}
 \partial_t U_k(\rho,\tau)&=&\frac{1}{2}\int_q \dot R_k(q^2)\Bigg[\frac{1}{Z_k {q}^2+R_k(q^2) + m_{1+}^{\ 2}}+ 
 \frac{1}{Z_k {q}^2+R_k(q^2)+m_{1-}^{\ 2}} 
 \nonumber
 \\
 \\
 \nonumber
&&\hspace{2cm} + \frac{1}{Z_k {q}^2+R_k(q^2) + m_{2+}^{\ 2}}+\frac{1}{Z_k {q}^2+R_k(q^2) + m_{2-}^{\ 2}} \nonumber \\
\nonumber \\
&&\hspace{2cm} +(N-2)\left(\frac{1}{Z_k {q}^2+R_k(q^2) + m_{3+}^{\ 2}}+\frac{1}{Z_k {q}^2+R_k(q^2) + m_{3-}^{\ 2}}\right)\Bigg]
\nonumber
\end{eqnarray}
where  the (momentum-dependent)  square ``masses''  are given by:
\begin{equation}
\left\{
\begin{array}{ll}
m_{1\pm}^{\ 2} &=  2 U^{(1,0)}_k+ 2 \rho U^{(2,0)}_k+ \rho  U^{(0,1)}_k +2\rho\tau U^{(0,2)}_k+8\tau U^{(1,1)}_k \\
\displaystyle & \hspace{0.5cm}  \pm   \left\{\tau\left(4 U^{(0,1)}_k + 4U^{(2,0)}_k + 4 \tau U^{(0,2)}_k + 4 \rho U^{(1,1)}_k \right)^2 
 +\left(\rho^2-4 \tau\right)\left(2 U^{(2,0)}_k-U^{(0,1)}_k
-2\tau U^{(0,2)}_k\right)^2 \right\}^{1\over 2}\\
\vspace{0cm} 
\displaystyle m_{2\pm}^{\ 2}& =  \displaystyle2 U^{(1,0)}_k+ \rho  U^{(0,1)}_k + 
{\omega_k\over 4} \rho\,  {q}^2 \pm{1\over 2} \left\{\omega_k^2 \tau {q}^4 +\left(\rho^2-4 \tau\right)\left(-{\omega_k\over 2} {q}^2+2  U^{(0,1)}_k \right)^2 \right\}^{1\over 2}\\
\\
\displaystyle m_{3\pm}^{\ 2} &= 2 U^{(1,0)}_k\pm 2\sqrt{\tau} U^{(0,1)}_k
\label{masses}
\end{array}
\right.
\end{equation}
\end{widetext}
with    $U^{(m,n)}_k=\partial^{m+n}  U_k(\rho,\tau)/\partial^m \rho\, \partial^n \tau$,  
$\rho={ \phi_1}^{\, 2}+{\phi_2}^{\, 2}$ and $\tau=({ \phi_1}^{\, 2}-{ \phi_2}^{\, 2})^2$ 
where $\phi_1$ and $\phi_2$ parametrize the configuration Eq.(\ref{confgeneric}).

We are interested, in the following, in fixed points of the RG flow equations. Finding them requires to work with dimensionless renormalized
quantities that are defined by: $\tilde\rho=Z_k k^{2-d}\rho$, $\tilde\tau=Z^2_k
k^{2(2-d)}\tau$,  $\widetilde{U}_k(\tilde\rho,\tilde\tau)=k^{-d}{U}_k(\rho,\tau)$,  
$\widetilde \omega_k=Z^{-2}_k k^{d-2} \omega_k$, $y=q^2/k^2$,  $r(y)=R_k(y k^2)/Z_k y k^2$.
In terms of these variables the flow equation of the potential writes:
\begin{eqnarray}
 \label{flotU}
\partial_t  \widetilde{U}_k\hspace{-0cm}&&=-d \widetilde{U}_k(\tilde\rho,\tilde\tau)+
(d-2+\eta_k)\Big[ \tilde{\rho} \widetilde{U}^{(1,0)}_k(\tilde\rho,\tilde\tau)\nonumber\\
&& \hspace{-0.cm}  +2 \tilde\tau \widetilde{U}^{(0,1)}_k(\tilde\rho,\tilde\tau)\Big]
+ 2 v_d \  \Big[l_{0}^d(\widetilde m_{1+}^{\ 2}) +  l_{0}^d(\widetilde m_{1-}^{\ 2})  \\
&&  \hspace{-0cm} + l_{0}^d(\widetilde m_{2+}^{\ 2}) 
 + l_{0}^d(\widetilde m_{2-}^{\ 2})  +  (N-2)\big(l_{0}^d(\widetilde m_{3+}^{\ 2}) +l_{0}^d(\widetilde m_{3-}^{\ 2})   \Big]\nonumber
\end{eqnarray}
where  $v_d^{-1}=2^{d+1}\pi^{d/2}\Gamma[d/2]$,  $\widetilde m_i$ are the dimensionless analogues of the masses
defined in Eqs.(\ref{masses}) and  the threshold functions $l_{n}^d(w)$  are defined by:
\begin{equation}
l_{n}^d(w)=-{n+\delta_{n,0}\over 2}\int_0^\infty dy\; y^{d/2}
\frac{\eta_k r(y) +2yr'(y)}{[y(1+r(y)) +w]^{n+1}} .
\end{equation}

The RG equations  for the  running anomalous  dimension $\eta_k$ and  the coupling constant $\omega_k$  
are given in terms of dimensionless quantities by:
\begin{widetext}
 \begin{eqnarray}
 \label{eta_frustre} 
\displaystyle  \eta_k &=& \displaystyle  \frac{2 v_d}{d} \Big[d \tilde\omega_k\,   {l^{d}_{1,0}}  \left(\widetilde m_{3}^{\ 2},0,0 \right) + 64 \tilde\kappa\,   (\widetilde U_k^{(0,1)})^2    m^d_{2,2} \left(\widetilde m_{3}^{\ 2},\widetilde m_{1-}^{\ 2},0\right) + 128 \tilde\kappa\, (\widetilde U_k^{(2,0)})^2    m^d_{2,2} \left(\widetilde m_{3}^{\ 2},\widetilde m_{1+}^{\ 2},0\right)  \nonumber  \\
&&+(d-2) \tilde\kappa \tilde\omega_k^2{l^{d+2}_{1,1}}  \left(\widetilde m_{3}^{\ 2},\widetilde m_{3}^{\ 2},\tilde\kappa\tilde\omega_k \right) +2 \tilde\kappa \tilde\omega_k^2m^{d+4}_{2,2} \left(\widetilde m_{3}^{\ 2},\widetilde m_{3}^{\ 2},\tilde\kappa\tilde\omega_k\right) -2\tilde\kappa  \tilde\omega_k^2 n^{d+2}_{1,2} \left(\widetilde m_{3}^{\ 2},\widetilde m_{3}^{\ 2},\tilde\kappa\tilde\omega_k\right)  \\
&&+  2\tilde\kappa   \tilde\omega_k^2 n^{d+2}_{2,1} \left(\widetilde m_{3}^{\ 2},\widetilde m_{3}^{\ 2},\tilde\kappa\tilde\omega_k\right) -   2 \tilde\kappa^2 \tilde\omega_k^3 l^{d+4}_{1,2} \left(\widetilde m_{3}^{\ 2},\widetilde m_{3}^{\ 2},\tilde\kappa\tilde\omega_k\right) + 2 \tilde\kappa^2 \tilde\omega_k^3 n^{d+4}_{2,2} \left(\widetilde m_{3}^{\ 2},\widetilde m_{3}^{\ 2},\tilde\kappa\tilde\omega_k\right)\Big] \nonumber
\end{eqnarray}
\end{widetext}

\begin{widetext}
 \begin{eqnarray}
 \label{omega_frustre}
\displaystyle  \frac{d\tilde\omega_k}{ dt}&=&\displaystyle (d-2+ 2 \eta)\, \tilde\omega_k + \frac{2 v_d}{d}
\Bigg[{d\tilde\omega_k\over \tilde\kappa} \big( l_{1,0}^d(\widetilde m_{3}^{\ 2},0,0)- l_{1,0}^d(\widetilde m_{1+}^{\ 2},0,0)\big)   +  (d-2) \tilde\omega_k^2 \,l_{1,1}^{d+2}(\widetilde m_{3}^{\ 2}, \widetilde m_{3}^{\ 2},\tilde\kappa\tilde\omega_k) \nonumber\\
&+& \hspace{-0.0cm}(10-d)\,\tilde\omega_k^2 \,l_{1,1}^{d+2}(\widetilde m_{1+}^{\ 2}, \widetilde m_{3}^{\ 2},\tilde\kappa\tilde\omega_k)-2\tilde\kappa\tilde\omega_k^3 \,l_{1,2}^{d+4}(\widetilde m_{3}^{\ 2}, \widetilde m_{3}^{\ 2},\tilde\kappa\tilde\omega_k)+6\tilde\kappa\tilde\omega_k^3 \,l_{1,2}^{d+4}(\widetilde m_{1+}^{\ 2}, \widetilde m_{3}^{\ 2},\tilde\kappa\tilde\omega_k)\\
&+&  \hspace{-0.0cm} 4(N-2)\,\tilde\omega_k^2 \,l_{2,0}^{d+2}(\widetilde m_{3}^{\ 2}, 0,0)+ 4  \tilde\omega_k^2 \,l_{2,0}^{d+2}(\widetilde m_{1-}^{\ 2}, 0,0)+ 64  (\widetilde U_k^{(0,1)})^2    m^d_{2,2} \left(\widetilde m_{3}^{\ 2},\widetilde m_{1-}^{\ 2},0\right) +2 \tilde\omega_k^2  m^{d+4}_{2,2} \left(\widetilde m_{3}^{\ 2},\widetilde m_{3}^{\ 2},\tilde\kappa\tilde\omega_k\right)  \nonumber \\
&-&  \hspace{-0.0cm}  2 \tilde\omega_k^2  m^{d+4}_{2,2} \left(\widetilde m_{1+}^{\ 2},\widetilde m_{3}^{\ 2},\tilde\kappa\tilde\omega_k\right)-  2 \tilde\omega_k^2  n^{d+2}_{1,2} \left(\widetilde m_{3}^{\ 2},\widetilde m_{3}^{\ 2},\tilde\kappa\tilde\omega_k\right)+6 \tilde\omega_k^2  n^{d+2}_{1,2} \left(\widetilde m_{1+}^{\ 2},\widetilde m_{3}^{\ 2},\tilde\kappa\tilde\omega_k\right)+ 2 \tilde\omega_k^2  n^{d+2}_{2,1} \left(\widetilde m_{3}^{\ 2},\widetilde m_{3}^{\ 2},\tilde\kappa\tilde\omega_k\right) \nonumber  \\
&-&  \hspace{-0.0cm}  6 \tilde\omega_k^2  n^{d+2}_{2,1} \left(\widetilde m_{1+}^{\ 2},\widetilde m_{3}^{\ 2},\tilde\kappa\tilde\omega_k\right)+  128 (\widetilde U_k^{(2,0)})^2    \left(m^d_{2,2} \left(\widetilde m_{3}^{\ 2},\widetilde m_{1+}^{\ 2},0\right) -m^d_{2,2} \left(\widetilde m_{1+}^{\ 2},\widetilde m_{3}^{\ 2},0\right)-\tilde\kappa\tilde\omega_k n^d_{2,2}\left(\widetilde m_{1+}^{\ 2},\widetilde m_{3}^{\ 2},\tilde\kappa\tilde\omega_k\right)\right)  \nonumber  \\
&+&  \hspace{-0.0cm}  8 \widetilde U_k^{(2,0)} \Big(d\, l_{1,1}^{d}(\widetilde m_{1+}^{\ 2}, \widetilde m_{3}^{\ 2},\tilde\kappa\tilde\omega_k)  -6 \tilde\kappa\tilde\omega_k l_{1,2}^{d+2}(\widetilde m_{1+}^{\ 2}, \widetilde m_{3}^{\ 2},\tilde\kappa\tilde\omega_k)+ 4 m^{d+2}_{2,2} \left(\widetilde m_{1+}^{\ 2},\widetilde m_{3}^{\ 2},\tilde\kappa\tilde\omega_k\right)- 6 n^{d}_{1,2} \left(\widetilde m_{1+}^{\ 2},\widetilde m_{3}^{\ 2},\tilde\kappa\tilde\omega_k\right) \nonumber \\
&+&  \hspace{-0.0cm}  6  n^{d+2}_{2,1} \left(\widetilde m_{1+}^{\ 2},\widetilde m_{3}^{\ 2},\tilde\kappa\tilde\omega_k\right)+4 \tilde\kappa\tilde\omega_k n^{d+2}_{2,2} \left(\widetilde m_{1+}^{\ 2},\widetilde m_{3}^{\ 2},\tilde\kappa\tilde\omega_k\right)\Big) + 2 \tilde\kappa\tilde\omega_k^3 n^{d+4}_{2,2} \left(\widetilde m_{3}^{\ 2},\widetilde m_{3}^{\ 2},\tilde\kappa\tilde\omega_k\right)-2 \tilde\kappa\tilde\omega_k^3  n^{d+4}_{2,2} \left(\widetilde m_{1+}^{\ 2},\widetilde m_{3}^{\ 2},\tilde\kappa\tilde\omega_k\right)  \Bigg] \nonumber 
 \hspace{-0.3cm}
\end{eqnarray}
\end{widetext}
where the masses are {evaluated in the configuration $\Phi_I$, Eq.(\ref{diagconf})  -- where $\tau=0$ --  and are  given by: 
\begin{equation}
\left\{
\begin{array}{ll}
& \widetilde m_{1+}^{\ 2} =  2 \widetilde U^{(1,0)}_{k}+ 8 \widetilde{\kappa} \widetilde U^{(2,0)}_{k}  \\
\\
& \widetilde m_{1-}^{\ 2}= \widetilde m_{2+}^{\ 2} =2 \widetilde U^{(1,0)}_{k}+ 4 \widetilde\kappa\widetilde U^{(0,1)}_{k} \\
\\
& \displaystyle\widetilde m_{3}^{\ 2}= \widetilde m_{3+}^{\ 2}   =\widetilde m_{3-}^{\ 2} = \widetilde m_{2-}^{\ 2}  =2\widetilde U^{(1,0)}_{k}
\label{massestau0}
\end{array}
\right.
\end{equation}
where  $U^{(a,b)}_{k}$ stands here for $U^{(a,b)}_{k}(\tilde\rho=2\widetilde{\kappa} ,\tau~=~0)$ and where the  threshold functions 
$l_{n_1,n_2}^d$, $m_{n_1,n_2}^d$  and $n_{n_1,n_2}^d$ are given  in the Appendix (\ref{threshold}).

\section{Field-semi-expansion}

The part of the potential that {\it a priori} needs to be represented  as accurately as possible is the vicinity 
of the minimum Eq.(\ref{groundstate}) since it describes the thermodynamics of the system.
The minimum occurs at a finite value $\rho$  and at vanishing $\tau$. We thus expect that the non trivial field dependence occurs
in the $\rho$\,-\,direction and not in the $\tau$ one. The idea of the field-semi-expansion is thus to keep the full $\rho$-dependence
of $U_k$ and to expand it in powers of  $\tau$: 
\be
\widetilde U_k(\tilde \rho,\tilde \tau)=\sum_{p=0}^{p_{\rm max}} \widetilde  U_{p,k}(\tilde \rho)\ \tilde \tau^p\  .
\label{expansion}
\ee
The flow  of  the  functions $\widetilde U_{p,k}(\tilde \rho)$ can be easily obtained by differentiating Eq.(\ref{flotU}) 
with respect to $\tau$.   We have  truncated the expansion at $p_{\rm max}=2, 3$ and 4.  
For the sake of simplicity we only display  the flow of  $ \widetilde{U}_{0,k}(\tilde\rho)$: 
\begin{eqnarray}
\label{eqpot0}
\partial_t  \widetilde{U}_{0,k}(\tilde \rho) &=&\hspace{-0cm}-d\, \widetilde{U}_{0,k}(\tilde \rho) 
+(d-2+\eta_k)\, \tilde{\rho}\,  \widetilde{U}'_{0,k}(\tilde \rho)
\nonumber\\
&&\hspace{-0.5cm}+2\,  v_d \left[ l_{1,0}^d (\widetilde m_{1+}^{\ 2},0 ,0)+2\, l_{1,0}^d(\widetilde m_{1-}^{\ 2},0,0) \right.\\
&&\left. \hspace{-0.5cm}+   l_{1,0}^d(0,\widetilde m_{1-}^{\ 2},\tilde \rho\tilde\omega_k/2)+2 (N-2)\,  \, l_{1,0}^d(\widetilde m_{3+}^{\ 2},0,0)\right]\nonumber
 \end{eqnarray}
with: $\widetilde{U}'_{0,k}(\tilde \rho)=d\widetilde{U}_{0,k}(\tilde \rho)/ d \tilde \rho$ and where the square masses are given by:
\begin{equation}
\left\{
\begin{array}{ll}
& \widetilde m_{1+}^{\ 2} =2\widetilde{U}'_{0,k}(\tilde \rho)+ 4 \tilde \rho  \widetilde{U}''_{0,k}(\tilde \rho)\\
\\
& \widetilde m_{1-}^{\ 2}=2\widetilde{U}'_{0,k}(\tilde \rho)+ 2 \tilde \rho  \widetilde {U}_{1,k}(\tilde \rho)\\
\\
& \displaystyle  \widetilde m_{3+}^{\ 2} =2\widetilde {U}'_{0,k}(\tilde \rho)\ . 
\end{array}
\right.
\end{equation}

\subsection{Procedure} 
In order to integrate the RG flow equations of  the functions $\widetilde{U}_{p,k}(\tilde\rho)$, we need an initial condition, that is, 
their values at $k=\Lambda$. We choose the usual  Ginzburg-Landau-Wilson potential:
\begin{equation}
\widetilde{U}_{k=\Lambda}(\rho,\tau)=\widetilde{r}_{\Lambda}\,\widetilde{\rho } + \widetilde g_{1,\Lambda}\, \widetilde{\rho}^{\,2} 
+ \widetilde g_{2,\Lambda}\,\widetilde\tau 
\end{equation}
which implies that $\widetilde{U}_{0,k=\Lambda}(\rho)=\widetilde{r}_{\Lambda}\,\widetilde{\rho } + 
\widetilde g_{1,\Lambda} \widetilde{\rho}^{\,2} $,
$\widetilde{U}_{1,k=\Lambda}(\rho)= \widetilde g_{2,\Lambda}\widetilde\tau$ and $\widetilde{U}_{p>1,k=\Lambda}(\rho)=0$.
Criticality can be reached by varying the mass parameter $\widetilde r_{\Lambda}$ while keeping fixed 
$\widetilde g_{1,\Lambda}$ and $\widetilde g_{2,\Lambda}$.This can be repeated in principle in any dimension $d$ and for all 
values of $N$ greater than $N_c(d)$.  We focus on the three dimensional case  and thus on  the value  of $N_c(d=3)$. 
We recall that  $N_c(d)$ results  from the collapse of two fixed points, one stable -- the chiral $C_+$ -- and one unstable -- 
the anti-chiral $C_-$  -- when $N$ is decreased from $N>N_c(d)$ to $N_c(d)$. These two fixed points are  related by a  RG flow line.  
When  $C_+$  and $C_-$  get closer, the speed of the flow decreases  and vanishes at  $N=N_c(d)$. Thus,   a  way to  determine  
$N_c(d)$ consists in identifying the value of $N$ for which the  first correction to scaling  critical  exponent $\omega$  vanishes.  In order to compute this quantity, we parametrize   the  potential  close to  the fixed point  by:
\begin{equation}
\widetilde U_{k}(\widetilde\rho,\widetilde\tau)=\widetilde U^*(\widetilde\rho,\widetilde\tau)+ 
\widetilde F(\widetilde\rho,\widetilde\tau)\,  e^{-t/\nu} +  \widetilde G(\widetilde\rho,\widetilde\tau)\,  e^{\omega t} 
\end{equation} 
where $\nu$ is  the usual critical exponent associated with the relevant direction and $\omega$ 
the subleading critical  exponent.

\subsection{LPA} 
\label{LPA}

In order to measure the impact  of the derivative terms $\eta_k$ and $\omega_k$ on $N_c(d=3)$ we perform two calculations, one where
these two coupling constants  are set to 0 (LPA) and the other one where we take them into account (LPA').
As for  the potential we  consider,  in the expansion Eq.(\ref{expansion}), 
the functions   $\widetilde{U}_{p,k}(\tilde\rho)$ up to $p_{\rm max}=4$.  This allows us  to analyze the convergence of the 
expansion in   powers of $\tau$.  Finally we have {\it optimized}  our results.  Indeed, as well known,  
finite truncations of the effective action   in the field  and/or derivatives of the field  induce  a residual 
dependence of the physical quantities  on the  regulator $R_k$. 
By definition, an optimal regulator is such that the dependence of the physical quantity under study upon
this regulator is minimal: this is the Principle of Minimal Sensitivity (PMS). Applied to the O$(N)$ models, this principle  indeed leads
to the best determination of the critical exponents in $d=3$ in the sense that the results thus obtained are  closest
to the Monte Carlo values \cite{canet03a,canet03b}.  To optimize our  results,  we have  used  the exponential 
cut-off  Eq.(\ref{cutoffexp}) that we have extended  
to a whole family  of cut-off  functions parametrized  by a real number $\alpha$:
\begin{equation}
R_k^{\alpha}(q^2)=\alpha { Z_k\,{q}^2\over e^{{q}^2/k^2}-1}\ . 
\label{cutoffexpbis}
\end{equation}
We have varied $\alpha$ in  order to reach a point of minimal sensitivity, {\it i.e.} a point where the physical quantities -- 
here mainly $N_c(d)$, now denoted  $N_c(d,\alpha)$ -- are as  insensitive  as possible  to  $\alpha$. This obviously corresponds 
to an extremum of $N_c(d,\alpha)$ as a function of $\alpha$.

\subsubsection{The N=6 case}

\begin{figure}[tp]
{\includegraphics[width=0.4\textwidth]{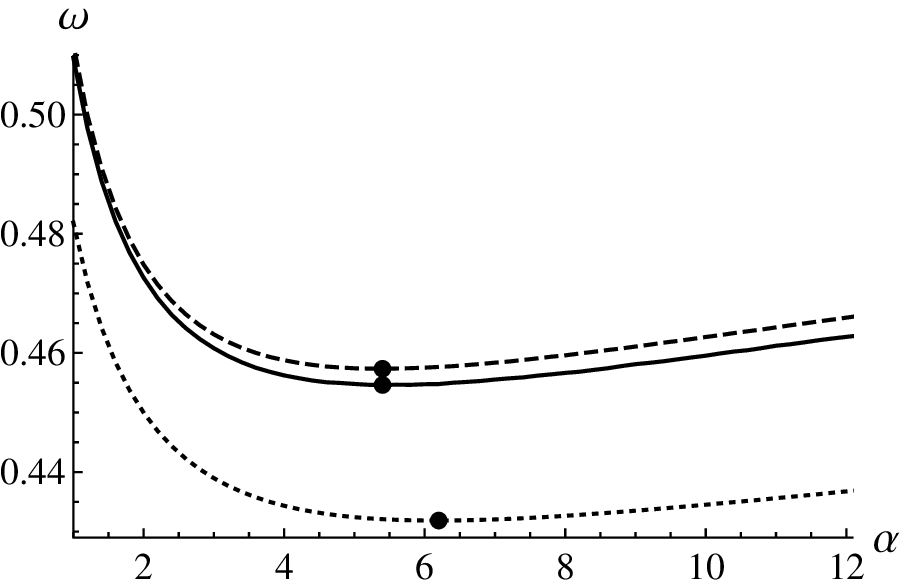}}\hspace{-0.5cm}{\includegraphics[width=0.4\textwidth]{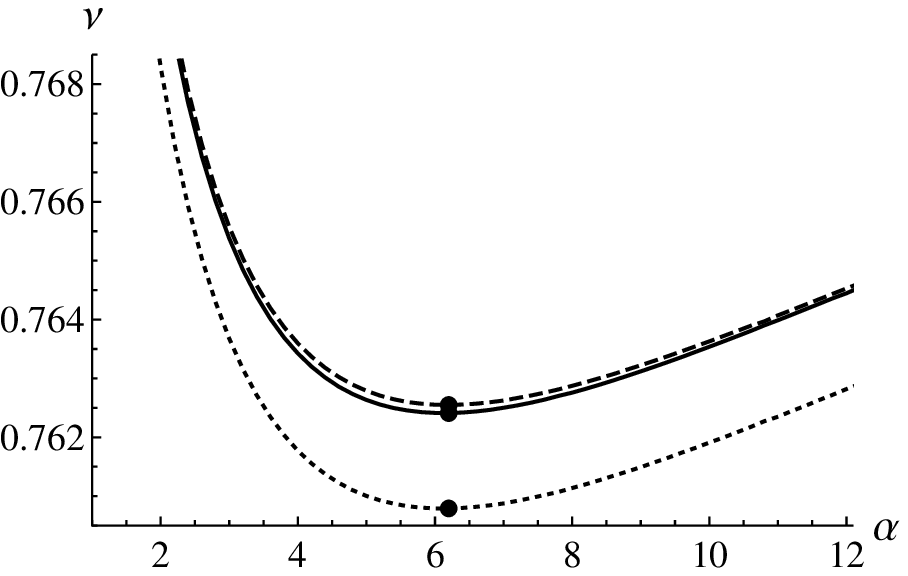}}
\caption{$N=6$, $d=3$ case:  critical exponents  $\omega$  (top) 
and  $\nu$ (bottom)   as functions of  the  regulator parameter $\alpha$ (see Eq.(\ref{cutoffexpbis})) calculated  
with  $p_{\rm max}=2$ (dotted curve), 3 (dashed curve) and 4 (solid curve). The black dots  indicate 
the position of the minima  of the curves.}
\label{N6}
\end{figure}

Before discussing the value of $N_c(d)$ within the derivative expansion, it is useful   to consider  a value of $N$ 
 where, very likely,    there exists a fixed point $C_+$, {\it i.e.} $N>N_c(d=3)$.  We  choose $N=6$ since 
a clear second order phase  transition has been found in this case by Monte Carlo simulations} \cite{loison00}.  
This allows us to check the convergence of our computation.  We find a 
stable fixed point  for $N=6$ in agreement with  the results obtained numerically \cite{loison00}.  Figure (\ref{N6}) 
displays the correction-to-scaling critical exponent  $\omega$ and the correlation length critical exponent $\nu$ as 
functions of  the regulator parameter 
$\alpha$  with  $p_{\rm max}=2,3$ and 4.  First, one  finds, for any  $p_{\rm max}$,  a unique extremum  for  each  
curve  $\omega(d=3,\alpha)$ and $\nu(d=3,\alpha)$ when varying $\alpha$.  Second,  one clearly observes a  
very good  convergence with $p_{\rm max}$. This means that the 
PMS can be safely applied and that   optimal  values  of the critical exponents can be determined.  
We find for $p_{\rm max}=4$:  $\omega_{\rm opt.}=\omega(d=3,\alpha=5.4)=0.455(5)$, $\nu_{\rm opt.}=\nu(d=3,\alpha=6.2)=0.7625(5)$ where error bars are evaluated  from the difference between two successive orders of the field expansion.  The value of $\nu$ can be compared with   the Monte Carlo value  $\nu=0.700(11)$ \cite{loison00}. Clearly our value of $\nu$ lies well above  the numerical result. This relies on the fact that the  effects of derivative terms have been neglected -- see below.

\subsubsection{\bf $N_c(d=3)$}

 \begin{figure}[tp]
{\includegraphics[width=0.35\textwidth]{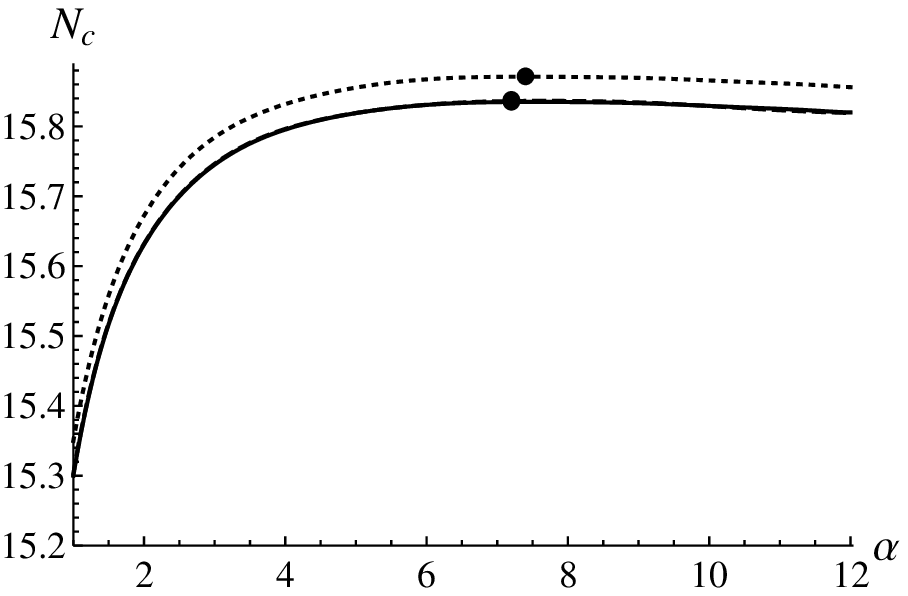}}
{\includegraphics[width=0.35\textwidth]{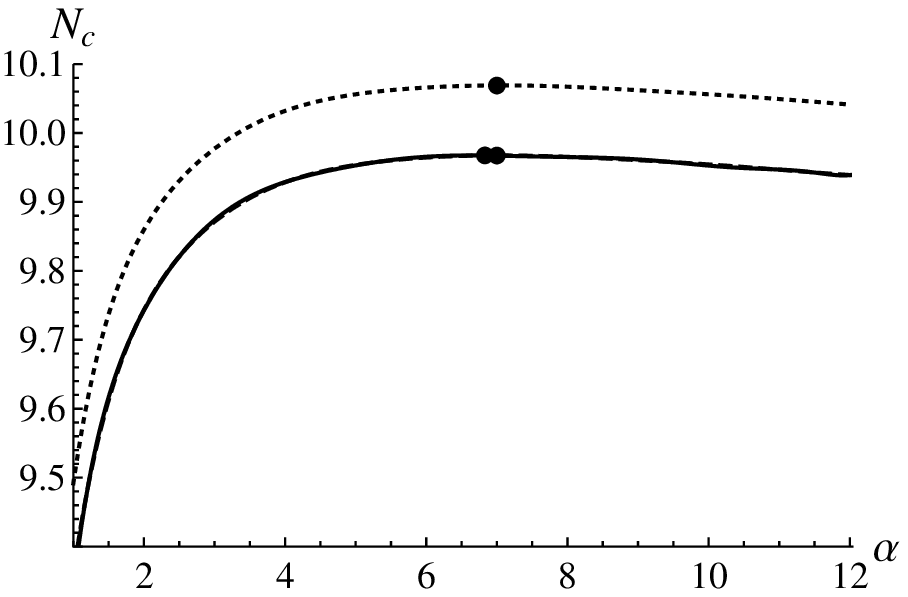}}
{\includegraphics[width=0.35\textwidth]{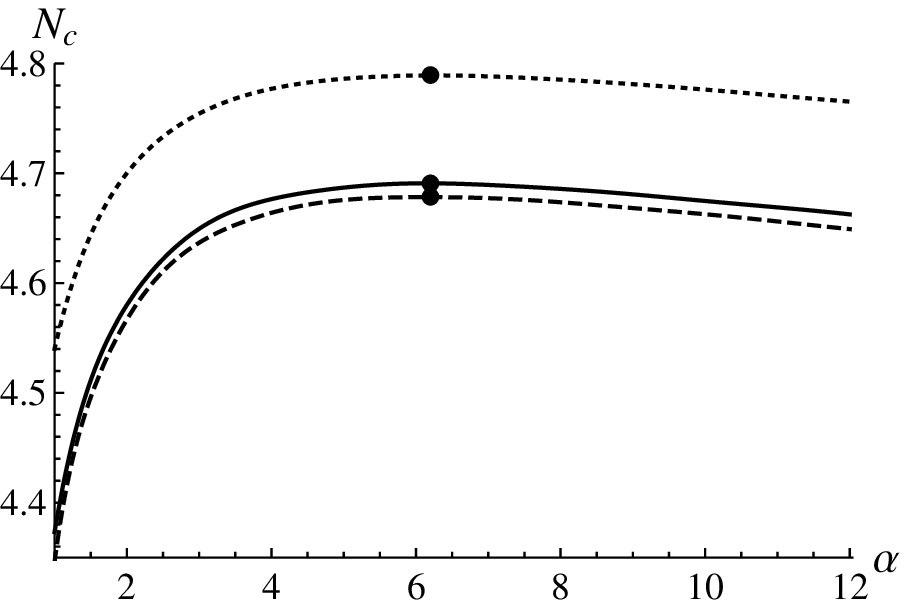}}
{\includegraphics[width=0.35\textwidth]{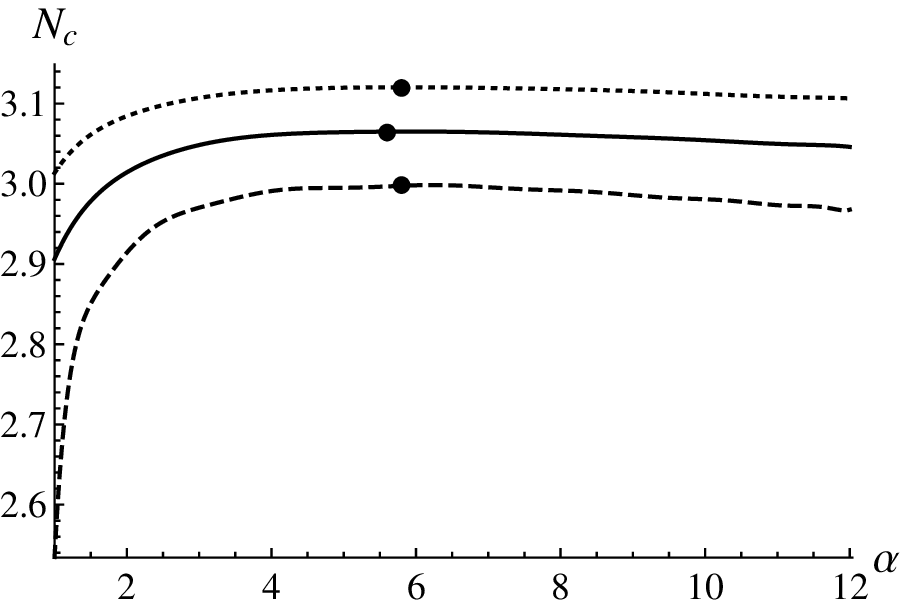}}
\caption{$N_c(d,\alpha)$  obtained with the  LPA with  $p_{\rm max}=2$  (dotted curve), $p_{\rm max}=3$ (dashed curve) 
and $p_{\rm max}=4$ (solid curve) for,  from top to bottom: $d=3.8$, $d=3.5$, $d=3$  and $d=2.8$. For $d=3.8$ and $d=3.5$, the two curves
obtained with $p_{\rm max}=3$ and $p_{\rm max}=4$ are superimposed at this  scale. The black dots  indicate 
the position of the minima  of the curves.}
\label{nch}
\end{figure}

We have then computed $N_c(d,\alpha)$ for $d=3.8, 3.5, 3.0$ and 2.7 within the LPA and for $p_{\rm max}=2,3,4$. In all cases,
we find a maximum  of $N_c(d,\alpha)$ when varying $\alpha$,  see Fig.(\ref{nch}).  Moreover we find that $N_c(d,\alpha)$ at its maximum  
converges when $p_{\rm max}$ is increased. However,  it appears that the speed of convergence  decreases   drastically  with  the dimension 
$d$, see Fig.(\ref{nch}).   The convergence is very  good   for $3<d<4$ and  becomes bad  typically around $d=2.8$.   This is clearly a limit of the field expansion performed here. Indeed    standard  power counting implies  the relevance of   more and more  powers of the field as the dimension is decreased. Another  problem  encountered at low dimensions is that there exists, for any dimension $d$ lower that $d=3$, a line $\widetilde N(d)$ located above the line $N_c(d)$ where   the critical exponent $\omega$  vanishes; this is again an artifact of  the field expansion. For these reasons  we  focus, in this article,  on  dimensions  between  $d=2.8$ and $d=4$.  In the physically relevant  $d=3$ case  one finds  that the  optimal  value of $N_c(d=3,\alpha)$ is given by $N_{c, \rm opt.}=N_c(d=3,\alpha=6.2)=4.68(2)$.  This value is almost  identical  to that found by  Zumbach    
\cite{zumbach93,zumbach94,zumbach94c} who has found $ N_c(d=3)=4.7$  using the  LPA  implemented  on the Polchinski equation  {\it without} any field expansion.  This suggests  that our result   for $N_c(d=3)$  is almost converged as far the field expansion  is concerned. This will be  confirmed  in section \ref{fullpotential} where we evaluate $N_c(d=3)$ without any field expansion.

\subsection{LPA' } 

Let us now consider   the contributions coming from   the leading derivative terms and thus  the impact of the 
anomalous dimension $\eta_k$ and of the coupling constant $\omega_k$ onto   $N_c(d)$.   The corresponding flow equations are 
given by Eqs.(\ref{eta_frustre})  and (\ref{omega_frustre}).   The  
flow of these new coupling constants  involves  a new  degree of freedom which is the choice  of  the field configuration $\Phi_I$
where  they are evaluated (see Eqs.(\ref{defZ}), (\ref{defomega}) and (\ref{diagconf})). In order  
 to implement the PMS we vary $\Phi_I$ and look for a extremum of the quantities we compute, that is,  the critical  exponents for  $N=7$, $N=6$ and  finally $N_c(d)$.

\subsubsection{The N=7 case}

\begin{figure}[tp]
{\includegraphics[width=0.4\textwidth]{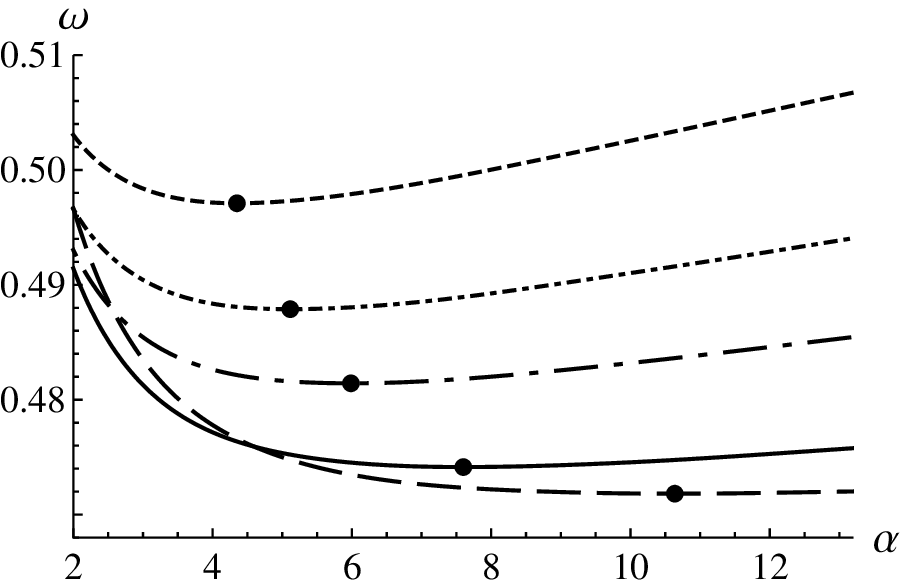}}\hspace{1cm}{\includegraphics[width=0.4\textwidth]{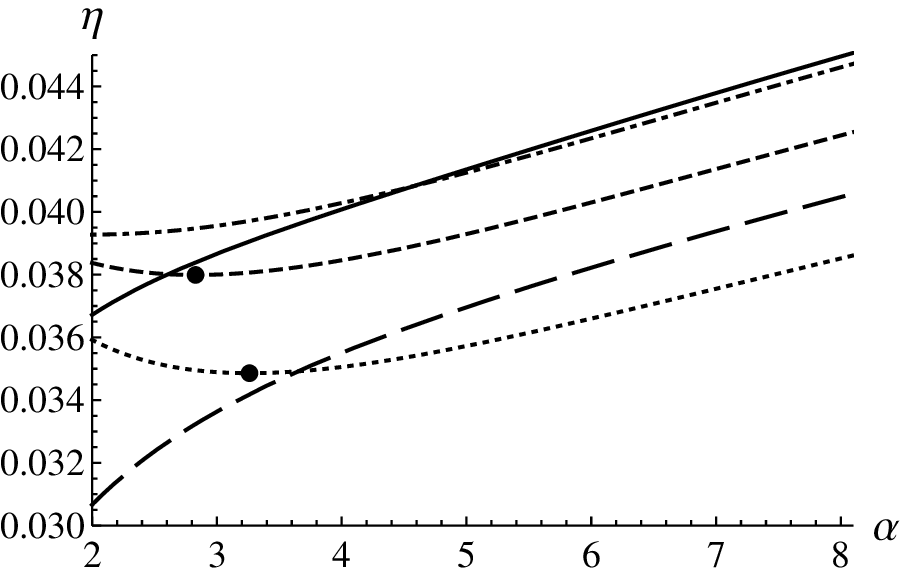}}
\caption{Critical exponents  $\omega$ (top) and $\eta$  (bottom)  as functions of the parameter 
$\alpha$ for $N=7$, $d=3$, $p_{\rm max}=4$.  Curves of different styles  correspond to various values  of $\rho_{\rm fix}$:  the dotted curve  corresponds to $\rho_{\rm fix}$ =0.4, the dashed curve  to $\rho_{\rm fix}$ =0.5, the dotdashed curve to $\rho_{\rm fix}$ =0.6, the long dotdashed 
curve  to $\rho_{\rm fix}$ =0.7, the solid curve to $\rho_{\rm fix}$ =0.9 and the long dashed curve to $\rho_{\rm fix}$=1.2. The black dots  indicate 
the position of the minima  of the curves.}
\label{omegan7prime} 
\end{figure}

\begin{figure}[tp]
{\includegraphics[width=0.4\textwidth]{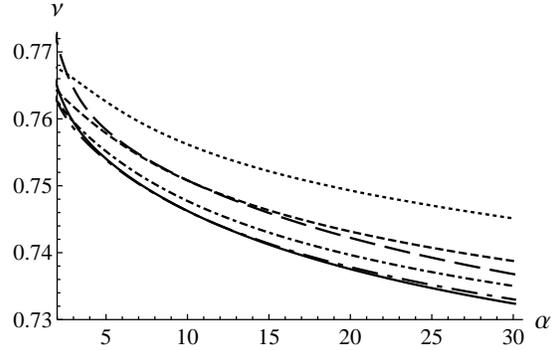}}
\caption{Critical exponent $\nu$ as a function of the parameter $\alpha$ for $N=7$, $d=3$, $p_{\rm max}=4$.  
The dotted curve  corresponds to $\rho_{\rm fix}$ =0.4, the dashed curve to $\rho_{\rm fix}$ =0.5,  the dotdashed curve  to 
$\rho_{\rm fix}$ =0.6, the long dotdashed curve to $\rho_{\rm fix}$ =0.7, the solid curve to $\rho_{\rm fix}$ =0.9
and the long dashed curve to $\rho_{\rm fix}$=1.2. There are no extrema for the  different  curves. }
\label{nun7prime}
\end{figure}

We first address   the $N=7$ case which  corresponds to the  smallest integer value of $N$ for which there is a stable fixed point both within the NPRG and  the perturbative  approaches  performed either within the $\epsilon$- or  pseudo-$\epsilon$- expansion.  We  find a fixed point for all $p_{\rm max}$.  The curves for the critical  exponents $\omega$, $\eta$  and $\nu$ as functions of $\alpha$ computed with $p_{\rm max}=4$ in $d=3$ 
are given in  Fig.(\ref{omegan7prime}) and Fig.(\ref{nun7prime}).   The different curves correspond to 
different values of  $\Phi_I$ or, equivalently, to different values of $\rho_{\rm fix}\equiv 2\kappa$,  
going from  0.4 to 1.2.  Although calculations were performed for all $\rho_{\rm fix}$  between 0.4 to 1.2 with step 0.1 we present only  the main curves in order not to overload  the figures. The results for $\omega$ show  that  a stationary curve is obtained for 
$\rho_{\rm fix}\simeq 0.9$  (solid  curve  in Fig.(\ref{omegan7prime})). For the corresponding 
curve, the  minimum is reached for   $\alpha\simeq 7.5$ which provides the optimal value  
$\omega_{\rm opt.}=\omega(d=3,\alpha=7.5)\simeq 0.475$. For $\eta$, the  stationary curve is obtained 
for  $\rho_{\rm fix}\simeq 0.6$ (dotdashed curve in Fig.(\ref{omegan7prime})). In this case 
however there is no genuine stationarity in $\alpha$. By continuity with the results  obtained for $\rho_{\rm fix}\simeq 0.4$ and $\rho_{\rm fix}\simeq 0.5$ (respectively dotted and dashed curves  on Fig.(\ref{omegan7prime})) for which we get genuine minima one can consider that there exists,   in the case  $\rho_{\rm fix}\simeq 0.6$,  a quasi-minimum  at   $\alpha=2$.  This provides the   almost optimal value $\eta_{\rm opt.}=\eta(d=3,\alpha=2)=0.039$. The determination of $\nu$  is more problematic, see Fig.(\ref{nun7prime}).  Indeed, whereas there is a stationary  curve when varying  $\rho_{\rm fix}$ at  $\rho_{\rm fix}=0.7$ (long dotdashed  curve  in Fig.(\ref{nun7prime}))  
there is no extremum when varying $\alpha$. However as the curve $\nu(d=3,\alpha)$  varies only  
weakly with $\alpha$  for large values of $\alpha$ one can provide a rough  estimate of  $\nu(d=3,\alpha)$  by  
the range $[0.730-0.740]$. We finally provide the   values of the critical exponents with estimation of error bars: $\omega=0.475(2)$, $\eta=0.039(2)$ and $\nu=0.735(5)$. These values  can be compared with the those  obtained perturbatively. The  $\epsilon$-expansion performed at five loops  within  the $\overline{MS}$ scheme \cite{calabrese03c}  provides  $\omega=0.33(10)$, $\gamma=1.39(6)$ and $\nu=0.71(4)$ that leads to, through scaling relations,  $\eta=0.042(4)$. Computations, still at five-loop order   and  within  the $\overline{MS}$ scheme {\it without}Ê $\epsilon$-expansion \cite{calabrese04} leads to  $\omega=0.5(2)$, $\eta=0.047(15)$ and $\nu=0.68(4)$. Finally  within a six-loop  computation performed   using  the zero momentum massive scheme  and resummed using  the  conformal mapping technique  (Pad\'e approximant techniques provide close results)  one finds \cite{calabrese03b} : $\omega=0.23(5)$, $\eta=0.042(2)$ and $\nu=0.68(2)$. Being given the  large   error bars provided by both the perturbative and non-perturbative computations the results are all compatible so that the $N=7$ does not show strong indications of a disagreement between the different approaches.

\subsubsection{The N=6 case} 

\begin{figure}[tp]
{\includegraphics[width=0.4\textwidth]{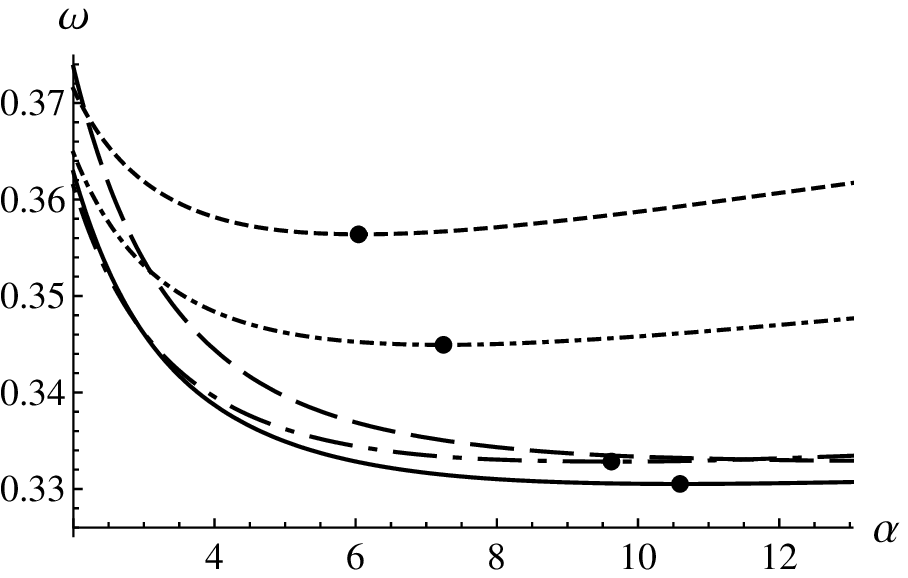}}\hspace{1cm}{\includegraphics[width=0.4\textwidth]{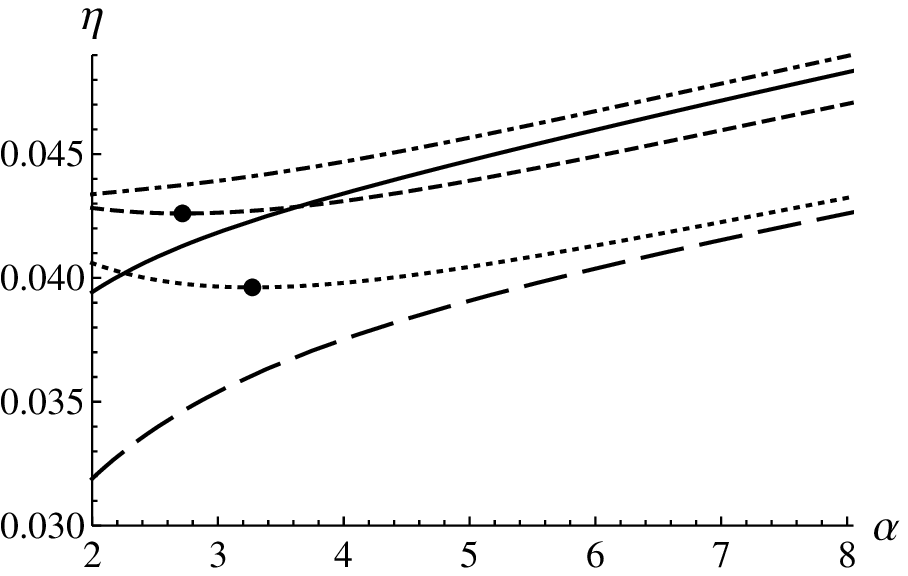}}
\caption{Critical exponents  $\omega$ (top) and $\eta$  (bottom)  as functions of the parameter 
$\alpha$ for $N=6$, $d=3$, $p_{\rm max}=4$.  Curves of different styles  correspond to various values  of $\rho_{\rm fix}$: the dotted curve  corresponds to $\rho_{\rm fix}$ =0.4, the
dashed curve  to $\rho_{\rm fix}$ =0.5, the dotdashed curve to $\rho_{\rm fix}$ =0.6, the long dotdashed 
curve  to $\rho_{\rm fix}$ =0.8, the solid curve to $\rho_{\rm fix}$ =0.9 and the long dashed curve to $\rho_{\rm fix}$=1.2.The black dots  indicate 
the position of the minima  of the curves.}
\label{omegan6prime} 
\end{figure}

\begin{figure}[tp]
{\includegraphics[width=0.4\textwidth]{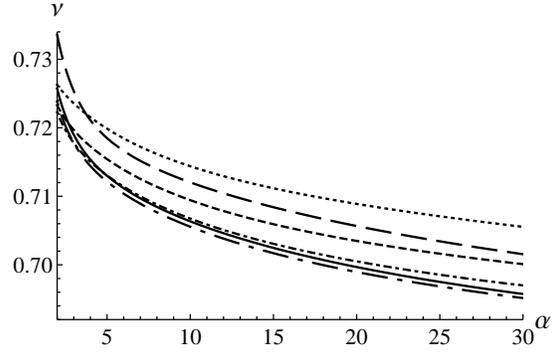}}
\caption{Critical exponent $\nu$ as a function of the parameter $\alpha$ for $N=6$, $d=3$, $p_{\rm max}=4$.  
The dotted curve  corresponds to $\rho_{\rm fix}$ =0.4, the dashed curve to $\rho_{\rm fix}$ =0.5,  the dotdashed curve  to 
$\rho_{\rm fix}$ =0.6, the long dotdashed curve to $\rho_{\rm fix}$ =0.8, the solid curve to $\rho_{\rm fix}$ =0.9 
and the long dashed curve to $\rho_{\rm fix}$=1.2. There are no extrema for the different curves. }
\label{nun6prime}
\end{figure}

For $N=6$, we  again find a fixed point for all $p_{\rm max}$.  The 
curves for the critical  exponents $\omega$, $\eta$  and $\nu$ as functions of $\alpha$ computed with $p_{\rm max}=4$ in $d=3$ 
are given in  Fig.(\ref{omegan6prime}) and Fig.(\ref{nun6prime}).  For $\omega$   a stationary curve is obtained for  $\rho_{\rm fix}\simeq 0.9$  (solid curve in Fig.(\ref{omegan6prime})) with a  minimum  reached for   $\alpha\simeq 10.5$. This  provides the optimal value  
$\omega_{\rm opt.}=\omega(d=3,\alpha=10.5)\simeq 0.330$. For $\eta$, a   stationary curve is obtained 
for  $\rho_{\rm fix}\simeq 0.6$ (dotted-dashed curve in Fig.(\ref{omegan6prime})). In this case, 
again, there is no genuine stationarity in $\alpha$. Note nevertheless that  for  close values 
of $\rho_{\rm fix}$ (of order 0.4--0.5,  corresponding to dotted and dashed curves in Fig.(\ref{omegan6prime})) 
one gets  clear  minima.  By continuity this provides an  almost optimal value of   $\eta$ lying between 0.040 and 0.045. The same problem as in the $N=7$ case is encountered for  $\nu$ --  see Fig.(\ref{nun6prime}) -- since,  whereas there is a stationary  curve when varying  
$\rho_{\rm fix}$ at  $\rho_{\rm fix}=0.8$ (long dotdashed curve  in Fig.(\ref{nun6prime}))  
there is no extremum when varying $\alpha$. As in the $N=7$ case one can consider the variation of  $\nu(d=3,\alpha)$ as smooth 
with $\alpha$  for large values of $\alpha$ and an estimate of  $\nu(d=3,\alpha)$ can be given by  
the range $[0.69-0.70]$. We finally provide the   values of the critical exponents with estimation of error bars: $\omega=0.330(5)$, $\eta=0.042(2)$ and $\nu=0.695(5)$.  The critical exponent $\nu$  can be favorably compared  with the Monte Carlo results $\nu=0.700(11)$ \cite{loison00},  what provides indications of convergence of  our computations. One can compare  these  results  with those obtained using a  five-loop computation  and performed  within  the $\overline{MS}$ scheme  {\it without}Ê $\epsilon$-expansion \cite{calabrese04} for which a fixed point is found  for all values of $N$. In the $N=6$ case,   these computations  lead to  $\eta=0.052(14)$ and $\nu=0.66(4)$.  These values are compatible with both the NPRG  and  Monte Carlo results. However it is at the cost of  large error bars, that are very likely  consequences of the poor  convergence of the computations performed at fixed dimensions already   observed in  \cite{delamotte08,delamotte10,delamotte10b}.

\subsubsection{\bf $N_c(d=3)$}

\begin{figure}[tp]
{\includegraphics[width=0.4\textwidth]{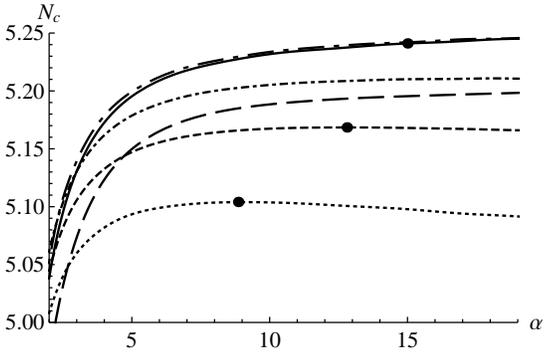}}
\caption{ $N_c(d=3,\alpha)$ for  $p_{\rm max}=4$. The dotted curve corresponds to $\rho_{\rm fix}$ =0.4, 
the dashed curve to $\rho_{\rm fix}$ =0.5, 
the dotdashed  curve   to $\rho_{\rm fix}$ =0.6, the long dotdashed curve  to $\rho_{\rm fix}$ =0.8, the solid curve  to 
$\rho_{\rm fix}$ =0.9 and the long dashed curve  to $\rho_{\rm fix}$=1.2. The black dots  indicate 
the position of the minima  of the curves.}
\label{ncfit} 
\end{figure}

Let us now consider $N_c(d=3)$ computed with $p_{\rm max}=4$  within the LPA'.  
We find an optimal value of  $N_c(d=3,\alpha)$  for   $\rho_{\rm fix}\in[0.8,0.9]$ (see solid and 
longdotdashed curves in  Fig.(\ref{ncfit})). However, there is no true extremum of $N_c(d=3)$ in the direction of $\alpha$ for  
these  values of  $\rho_{\rm fix}$ although $N_c(d=3)$ is  almost insensitive 
to $\alpha$  for  $\alpha \sim 15$, see Fig.(\ref{ncfit}).  This provides the best possible value of 
$N_c(d=3)$: $N_{c,{\rm opt}}(d=3,\alpha=15)=5.24(2)$.  The 
comparison with the value obtained using a usual  field expansion \cite{tissier00,tissier00b,tissier01, tissier03, delamotte03},   
$N_c(d=3)=5.1$,   shows  that the effects of  high orders in the field neglected in previous approaches   
were  not  negligible. Moreover,  compared  with the  value $N_c(d=3)=4.7$  obtained within  
the LPA, one sees that derivative terms  also play an important role.     Our  value of $N_c(d=3)$ can 
finally be  compared with those  obtained perturbatively   via the $\epsilon$- or pseudo-$\epsilon$-
expansions.  Within the $\epsilon$-expansion, one finds at five loops  $N_c(d = 3)=6.1(6)$ \cite{calabrese03c}, 
and the pseudo-$\epsilon$- expansion at six loops leads to $N_c(d = 3)=6.22(12)$  \cite{calabrese03c}  and 
$N_c(d = 3)=6.23(21)$ \cite{holovatch04}.   The quantitative agreement is  not excellent. 
One can suspect contributions of higher orders in  derivatives  of the field within the 
NPRG approach or/and  contributions of higher  orders in the loop expansions. However,  
from the qualitative point of view all these results agree as for  the absence of a non trivial fixed 
point for $N=2$ and $N=3$.

 \subsection{\bf The curve $N_c(d)$}

\begin{figure}[h]
{\includegraphics[width=0.4\textwidth]{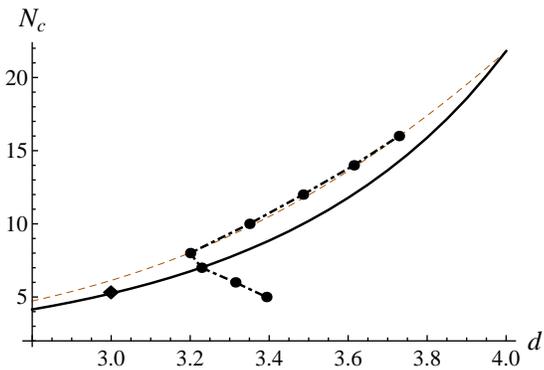}}
\caption{Curves $N_c(d)$. Solid curve  (this work):  LPA'  with four  functions. Diamond   (this work)  $N_c(d=3)$ obtained by  LPA'  with the full potential. Dashed curve :  perturbative  five loops ($\overline{MS}$ scheme)  with  $\epsilon$-expansion  \cite{calabrese03b}.  Dotdashed curve with black points: perturbative  six  loops ($\overline{MS}$ scheme)  without $\epsilon$-expansion  \cite{calabrese04}. }
\label{nnncritical} 
\end{figure}

Finally we have computed $N_c(d)$ for values of $d$ going from  $d=2.8$ to $d=4$.  The corresponding (solid) curve   is  shown  in Fig.(\ref{nnncritical}) together with that  (dashed)  obtained  within the perturbative five-loops  $\epsilon$-expansion  of   \cite{calabrese03b} and that (dotdashed) obtained within the perturbative six-loop    {\it without}  $\epsilon$-expansion \cite{calabrese04}. The qualitative agreement between the NPRG  and the $\epsilon$-expansion results  is  apparent although, quantitatively, there  remain some  relatively important gaps  between   the different values of  $N_c(d)$ for some dimensions between $d=3$ and $d=4$. On the other hand,  the singular character  of  the result obtained  {\it without}  $\epsilon$-expansion -- for which  the curve $N_c(d)$  displays a S-like shape --  is also obvious.  

\section{Full potential approach}
\label{fullpotential}

We finally  present  preliminary  results obtained  when  one  takes  into account  the full field content of the effective potential. In this case we have employed the theta  cut-off Eq.(\ref{cutoffstep}) what  allows to obtain analytical expressions for the RG equations and thus a manageable  
integration of the RG flow. However  due to  the massive computational  requirements of this approach   we have focused our attention on the $d=3$ case delaying the study of the general  case  in  \cite{delamotte15}.   Note that, for simplicity, we set  $\omega_k=0$ which is justified  by the observation that this coupling constant plays a minor role around $d=3$. First we have determined the critical exponent in the $N=7$ case and have obtained:   $\eta=0.0438$ and  $\nu=0.760$. These  values  are roughly  compatible with those obtained within the field-semi-expansion ($\eta=0.039(2)$ and $\nu=0.735(5)$). This is  also true in  the $N=6$ case for which one finds:  $\eta=0.0487$  and $\nu=0.716$ to be compared to  $\eta=0.042(2)$ and $\nu=0.695(5)$,  with a critical exponent  $\nu$ still  compatible with  that found within the Monte Carlo approach ($\nu=0.700(11)$).  As for  $N_c(d=3)$ the full potential approach leads to the value $N_c(d=3)=4.8$ using the LPA. Compared to the value   $N_c(d=3)=4.7$ obtained within the LPA in a field-semi-expansion this shows that our expansion was almost converged, as claimed  in section \ref{LPA}.  Taking now into account the derivative terms  at lowest order (LPA')  one finds: $N_c(d=3)=5.4$, see Fig.(\ref{nnncritical}). Compared with the value  $N_c(d=3)=5.24(2)$  obtained within the LPA' in a field-semi-expansion this confirms  that  the field expansion  is almost converged.  But compared to  the value $N_c(d=3)=4.8$ obtained  using the LPA this also shows that the effects of derivatives terms are  not negligible.  Clearly one cannot firmly conclude   that  convergence of the derivative expansion  has been reached  and one can expect that higher order terms  contribute. However   it is very unlikely that  these derivative terms  drastically change the overall shape of the curve  $N_c(d)$ obtained here.

\section{Conclusion}

We have investigated the behaviour of non-collinear magnets using  a functional RG approach, focusing 
on the critical value $N_c(d)$.  First, from the methodological point of view,  our approach, that  combines computations based on a field-semi-expansion and computations performed without any field expansion,  confirms the validity  of former approach  to investigate complex systems.  Second, as for the physics of frustrated magnets, our computations clearly favours  a  value  of  $N_c(d=3)$  significantly  larger   than 3 excluding the occurrence  of a  second order phase transition in the physical $N=2$ and $N=3$ cases.  Our  result  confirms  $\epsilon$- (and pseudo-$\epsilon$-) expansions  as well as early NPRG approaches  based on the Polchinski and  Wetterich equations.  It  contradicts  both those obtained using fixed dimension  perturbative approaches  as well as those   resulting from  recent CB approach.  Up to now there is no indication   of  failure  or even weakness  coming from either   $\epsilon$--expansion or NPRG approaches. On the opposite   the conclusion of second order phase 
transition obtained within fixed dimension approaches have been shown to be extremely  dubious. There is  no 
such  suspicion concerning the CB approach. However several points must be clarified in this context. First,  it 
is crucial to  understand the status of the assumptions made  and, in particular, that 
of the existence of scale invariance,  in order to avoid  a  self-fulfilling prophecy.  Second,  the method should give an account of the 
existence of the anti-chiral fixed point $C_-$ and the ``dynamics" of   $C_+$  and $C_-$ 
that  generates  the curve $N_c(d)$ when $N$ is lowered. 
 
 \acknowledgments
 
This work was supported in part by the 7th FP, IRSES project No 269139 ÒDynamics and Cooperative phenomena
in complex physical and biological environmentsÓ. M.D. thanks  the LPTMC for hospitality during preparation
of this work. 

\appendix
\section{Propagator}
\label{propagator}

The propagator of the model is a matrix  whose components are given  by  the (Fourier transform of the) 
second derivative of $\Gamma_k$ with respect to the field (see Eq.(\ref{secondderivatives})) evaluated in the  
configuration Eq.(\ref{confgeneric}). It is given by:

\begin{widetext}
\begin{equation}
\label{matrics}
\Gamma_{k,(a,i),(b,j)}^{(2)}[{q}_1,{q}_2]
+R_k({q})=\left(\begin{array}{ccccccccc}
\Gamma_{k,(1,1),(1,1)}&\Gamma_{k,(1,1),(1,2)}&\Gamma_{k,(1,1),(2,1)}&\Gamma_{k,(1,1),(2,2)}&&&&\\
\Gamma_{k,(1,2),(1,1)}&\Gamma_{k,(1,2),(1,2)}&\Gamma_{k,(1,2),(2,1)}&\Gamma_{k,(1,2),(2,2)}&&&&\\
\Gamma_{k,(2,1),(1,1)}&\Gamma_{k,(2,1),(1,2)}&\Gamma_{k,(2,1),(2,1)}&\Gamma_{k,(2,1),(2,2)}&&&&0&\\
\Gamma_{k,(2,2),(1,1)}&\Gamma_{k,(2,2),(1,2)}&\Gamma_{k,(2,2),(2,1)}&\Gamma_{k,(2,2),(2,2)}&&&&\\
&&&&&&\hspace{-0.7cm} \ddots&\\
&&&&&&&\Gamma_{k,(1,i>2),(1,i>2)}\\
&&&&&&&&\hspace{-1.cm} \Gamma_{k,(2,i>2),(2,i>2)}\\
&&0&&&&&&\hspace{-1.9cm}\ddots
 \end{array}\right),
\end{equation}
\end{widetext}
or, taking account of the vanishing components and the symmetries,  by:
\begin{equation}
\label{matrics2}
\Gamma_{k,(a,i),(b,j)}^{(2)}[{q}_1,{q}_2]
+R_k({q})=\left(\begin{array}{ccccccccccc}
 A&0&0&C&&&&\\
 0&E&D&0&&&&\\
 0&D&F&0&&&&0\\
 C&0&0&B&&&&\\
 &&&&& \hspace{-0.3cm} \ddots&\\
 &&&&&&H&&&&\\
 &&&&&&&G&&\\
 &&&&&&&&H&\\
 &&0&&&&&&&G\\
 &&&&&&&&&&\ddots
 \end{array}\right),
\end{equation}
where the matrix elements  $A, B, C, D, E, F, G$ and $H$, are given by:
\begin{widetext}
\begin{equation}
\begin{array}{ll}
A&=Z_k {q}^2+R_k(q^2)+2U_k^{(1,0)}+({ \phi_1}^{2}-{\phi_2}^{2})U_k^{(0,1)}+{\phi_1}^{2}\left(4U_k^{(2,0)}{+}4({ \phi_1}^{2}{-}{ \phi_2}^{2})U_k^{(1,1)}{+} 2U_k^{(0,1)}{+}({ \phi_1}^{2}-{ \phi_2}^{2})^2U_k^{(0,2)}\right)\\
\\
B&=Z_k {q}^2+R_k(q^2)+2U_k^{(1,0)}-({ \phi_1}^{2}-{ \phi_2}^{2})U_k^{(0,1)}+{ \phi_2}^{2}\left(4U_k^{(2,0)}{-}4({ \phi_1}^{2}{-}{ \phi_2}^{2})U_k^{(1,1)}{+}
2U_k^{(0,1)}{+}({ \phi_1}^{2}-{ \phi_2}^{2})^2U_k^{(0,2)}\right)
\end{array}
\end{equation}
\end{widetext}
\begin{equation}
\begin{array}{ll}
C&=\displaystyle \phi_1\phi_2\left(4U_k^{(2,0)}-2U_k^{(0,1)}-({\phi_1}^{2}-{\phi_2}^{2})^2U_k^{(0,2)}\right)\\
\\
D&=\displaystyle \phi_1\phi_2\left(-\frac{\omega_k}{2}{q}^2+2U_k^{(0,1)}\right)\\
\end{array}
\end{equation}

\begin{equation}
\begin{array}{ll}
E&=\displaystyle Z_k {q}^2+R_k(q^2)+\frac{\omega_k}{2}{ \phi_1}^{2}{q}^2+2U_k^{(1,0)}+\rho
U_k^{(0,1)}\\
\\
F&=\displaystyle Z_k {q}^2+R_k(q^2)+\frac{\omega_k}{2}{ \phi_2}^{2}{q}^2+2U_k^{(1,0)}+\rho U_k^{(0,1)}\\
\end{array}
\end{equation}

\begin{equation}
\begin{array}{ll}
G&=\displaystyle Z_k {q}^2+R_k(q^2)+2U_k^{(1,0)}-({ \phi_1}^{2}-{ \phi_2}^{\, 2})U_k^{(0,1)}\\
\\
H&=\displaystyle Z_k {q}^2+R_k(q^2)+2U_k^{(1,0)}+({\phi_1}^{2}-{\phi_2}^{2})U_k^{(0,1)}
\end{array}
\end{equation}

The eigenvalues of  Eq.(\ref{matrics2})  are given by:
\begin{equation}
\begin{array}{ll}
\lambda_{1,2}&= \frac{A+B\pm\sqrt{(A-B)^2+4C^2}}{2}\\
\\
\lambda_{3,4}&= \frac{E+F\pm\sqrt{(E-F)^2+4D^2}}{2}
\end{array}
\end{equation}
\begin{equation}
\lambda_{5}=\dots=\lambda_{2N-1}=H\qquad\lambda_{6}=\dots=\lambda_{2N}=G
\end{equation}

 Taking into account  of the fact that in  the  configuration Eq.(\ref{confgeneric})  that we consider 
 to establish the equation of the effective potential one has:
${\phi_1}^{2}+{\phi_2}^{2}=\rho$ and $({\phi_1}^{2}-{\phi_2}^{2})^2=2\tau$
 the eigenvalues read:
\begin{widetext}
\begin{equation}
\begin{array}{ll}
\lambda_{1\pm}&= Z_k {q}^2+R_k(q^2) + 2 U_k^{(1,0)}+ 2 \rho U_k^{(2,0)}+ \rho  U_k^{(0,1)} +\rho\tau U_k^{(0,2)}+8\tau U_k^{(1,1)} \\
\\
&\displaystyle \pm \left\{\tau\left(4 U_k^{(0,1)} + 4U_k^{(2,0)} + 4 \tau U_k^{(0,2)} + 4 \rho U_k^{(1,1)} \right)^2 +\left(\rho^2-4 \tau\right)\left(2 U_k^{(2,0)}-U_k^{(0,1)}
-2\tau U_k^{(0,2)}\right)^2 \right\}^{1\over 2}\\
\\
&=Z_k {q}^2+R_k(q^2) + m_{1\pm}^{\ 2}
\\
\\
\\
\displaystyle \lambda_{2\pm}&= \displaystyle Z_k {q}^2+R_k(q^2) + 2 U_k^{(1,0)}+ \rho  U_k^{(0,1)} + {\omega_k\over 4} \rho\,  {q}^2 \pm{1\over 2} \left\{\omega_k^2 \tau {q}^4 +\left(\rho^2-4 \tau\right)\left(-{\omega_k\over 2} {q}^2+2  U_k^{(0,1)} \right)^2 \right\}^{1\over 2}\\
\\
&=Z_k {q}^2+R_k(q^2) + m_{2\pm}^{\ 2}
\end{array}
\end{equation}
\end{widetext}
where one has to take care about the fact that  the "masses"   $m_{2+}$ and $m_{2-}$ are momentum-dependent.  There are moreover  $N-2$ modes with eigenvalues $\lambda_{3+}$ and $N-2$ modes with eigenvalues $\lambda_{3-}$ with:
\begin{equation}
\begin{array}{ll}
\displaystyle\lambda_{3\pm}& =Z_k {q}^2+R_k(q^2) + 2 U_k^{(1,0)}\pm 2\sqrt{\tau} U_k^{(0,1)} \\
\\
&=Z_k {q}^2+R_k(q^2) +m_{3\pm}^{\ 2}\ . 
\end{array}
\end{equation}

For completeness  the mass spectrum at the minimum of the potential is given. In this case one has:  $U_k^{(1,0)}=0$ and $\tau=0$. 
This implies that:
$m_{1+}^{\ 2}=4\rho U_k^{(2,0)}$, $m_{1-}^{\ 2}=2\rho U_k^{(0,1)}$, $m_{2+}^{\ 2}=2\rho U_k^{(0,1)}=m_{1-}^{\ 2}$ 
and $m_{2-}=m_{3+}=m_{3-}=0$. As a consequence one obtains the following spectrum:  one massive singlet with 
square mass  $m_s^2=4\rho U_k^{(2,0)}$, one  massive   doublet  with square  mass $m_d=2\rho U_k^{(0,1)}$ and  $2N-3$ 
Goldstone modes.

\vspace{0.5cm}

\section{The threshold functions}
\label{threshold}

We finally discuss the different threshold functions $l$, $m$
and $n$ appearing in the flow equations.  

\subsection{Definitions}

The threshold functions are defined as:
\begin{widetext}
\begin{equation}
\begin{array}{ll}
\displaystyle l_{n_1,n_2}^d(w_1,w_2,w)=
-\frac12\int_0^\infty dy\;&  y^{d/2-1}
\tilde{\partial}_t\displaystyle \left\{\frac{1}{(P_1+w_1)^{n_1}
(P_2+w_2)^{n_2}}\right\}\\
\\
\displaystyle m_{n_1,n_2}^d(w_1,w_2,w)=
-\frac12\int_0^\infty  dy\;& y^{d/2-1}
\tilde{\partial}_t \displaystyle \left\{\frac{y(\partial_y P_1)^2}{(P_1+w_1)^{n_1}
(P_2+w_2)^{n_2}}\right\}  \\
\\
\displaystyle n_{n_1,n_2}^d(w_1,w_2,w)=
-\frac12\int_0^\infty  dy \;&  y^{d/2-1}
\tilde{\partial}_t \displaystyle \left\{\frac{y\partial_y P_1}{(P_1+w_1)^{n_1}
(P_2+w_2)^{n_2}}\right\} 
\end{array}
\end{equation}
\label{annexe_def_thres}
where:
\begin{equation}
\left\{
\begin{array}{ll}
&P_1=P_1(y)=y(1+r(y))\\ 
\\
&P_2=P_2(y,w)=y(1+r(y)+w)
\end{array}
\right.
\end{equation}
\end{widetext}
and $r(y)$ is the  dimensionless cut-off:  $r(y)=R_k(y k^2)/Z_k y k^2$.

We recall that the tilde in $\tilde\partial_t$ means that only the $t$
dependence of the function $R_k$ is to be considered. As a
consequence, we should not consider the $t$-dependence of the coupling
constants to perform this derivative. Therefore, in the preceding
equations:
\begin{align}
\tilde{\partial}_t P_i =\frac{\partial R_k}{\partial t} 
\frac{\partial  P_i}{\partial R_k}=-y(\eta r(y)+2y r'(y)).
\end{align}
Now, threshold functions can be expressed as explicit integrals if we
compute the action  of $\tilde \partial_t$. To this end, it is
interesting to notice the equality: $\tilde{\partial}_t \partial_y P_i=\partial_y \tilde{\partial}_t P_i$, so that:
\begin{equation}
\tilde{\partial}_t\partial_y r(y)=-\eta\Big(r(y)+yr'(y)\Big)-2y\Big(2r'(y)+yr''(y)\Big)
\end{equation}
We then get:
\begin{widetext}
\begin{gather}
\begin{split}
\hspace{-3.5cm}l_{n_1,n_2}^d(w_1,w_2,w)=-\frac12\int_0^\infty dy\; y^{d/2}
\;&\frac{\eta r(y) +2yr'(y)}{(P_1+w_1)^{n_1}
(P_2+w_2)^{n_2}}\left(\frac{n_1}{P_1+w_1}+\frac{n_2}{P_2+w_2}
\right)\, \label{expression_pour_l}
\end{split}\\ \displaybreak[0]
\begin{split}
n_{n_1,n_2}^d(w_1,w_2,w)=-&\frac12\int_0^\infty dy\; y^{d/2}
\frac{1}
{(P_1+w_1)^{n_1}(P_2+w_2)^{n_2}} \Bigg\{y\Big(1+r(y)+yr'(y)\Big)\Big(\eta r(y)+2y
r'(y)\Big)\ \times  \\\  &\left(\frac{n_1}{P_1+w_1} +\frac{n_2}{P_2+w_2} 
\right)-\eta\Big(r(y)+yr'(y)\Big)-2y\Big(2r'(y)+yr''(y)\Big)
   \Bigg\}
\end{split}\\ \displaybreak[0]
\begin{split}
m_{n_1,n_2}^d(w_1,w_2,w)=-&\frac12\int_0^\infty dy\; y^{d/2}
\frac{1+r(y)+yr'(y)}
{(P_1+w_1)^{n_1}(P_2+w_2)^{n_2}}\Bigg\{y\Big(1+r(y)+yr'(y)\Big)\Big(\eta r(y)+2y
r'(y)\Big)\ \times  \\\ &\Bigg(\frac{n_1}{P_1+w_1}+\frac{n_2}{P_2+w_2}
\Bigg)-2\eta\Big(r(y)+yr'(y)\Big)-4y\Big(2r'(y)+yr''(y)\Big)\ \label{expression_pour_m}
   \Bigg\}\ .
\end{split}
\end{gather}
\end{widetext}

%



\end{document}